\documentclass[preprint]{ptephy_v1}%%%%%% to generate preprint number
\usepackage{subcaption}
\usepackage{caption}
\usepackage{multirow}
\usepackage{arydshln}
\usepackage{amssymb}
\usepackage{amsmath}
\usepackage{url} %It provides better support for handling and breaking URLs.
\usepackage{xfrac}
\usepackage{lscape}
\usepackage{epstopdf}

\newcommand{\GdSOw}{$\rm Gd_2(\rm SO_4)_3\cdot \rm 8H_2O$}
\newcommand{\GdSO}{$\rm Gd_2(\rm SO_4)_3$}

\newcommand{\GdOx}{$\rm Gd_2 \rm O_3$}

\begin{document}

\title{Development of Ultra-pure Gadolinium Sulfate for the Super-Kamiokande Gadolinium Project}

%%%% To generate auto affiliation numbers please use \author{}\affil{} command

\newcommand{\AFFicrr}{1}
\newcommand{\AFFmad}{2}
\newcommand{\AFFlsc}{3}
\newcommand{\AFFkek}{4}
\newcommand{\AFFokayama}{5}
\newcommand{\AFFsheff}{6}
\newcommand{\AFFbul}{7}
\newcommand{\AFFtohoku}{8}
\newcommand{\AFFtodai}{9}
\newcommand{\AFFipmu}{10}
\newcommand{\AFFuci}{11}
\newcommand{\AFFtus}{12}
\newcommand{\AFFtsukuba}{13}
\newcommand{\AFFnyc}{14}
\newcommand{\AFFpablonow}{*}
\newcommand{\AFFpereznow}{**}

\author[ ]{
%%%%%%%%%%%%%%%%%%%%%%%%%%%%%%%%%%%%%%%%%%%%%%%%%%%%%%%%%%%%%%%%%%%%
%ICRR
\name{K.~Hosokawa}{\AFFicrr},
\name{M.~Ikeda}{\AFFicrr},
\name{T.~Okada}{\AFFicrr},
\name{H.~Sekiya}{\AFFicrr,\AFFipmu},
%%%%%%%%%%%%%%%%%%%%%%%%%%%%%%%%%%%%%%%%%%%%%%%%%%%%%%%%%%%%%%%%%%%%%
%% Madrid
\name{P.~Fern\'andez}{\AFFmad}\thanks{currently at Donostia International Physics Center DIPC, San Sebasti\'an / Donostia, E-20018, Spain},
\name{L.~Labarga}{\AFFmad},
\name{I.~Bandac}{\AFFlsc},
\name{J.~Perez}{\AFFlsc}\thanks{currently at M. Smoluchowski Institute of Physics, Jagiellonian University, 30-348 Kraków, Poland},
%%%%%%%%%%%%%%%%%%%%%%%%%%%%%%%%%%%%%%%%%%%%%%%%%%%%%%%%%%%%%%%%%%%%%
%%KEK
\name{S.~Ito}{\AFFkek},
%%%%%%%%%%%%%%%%%%%%%%%%%%%%%%%%%%%%%%%%%%%%%%%%%%%%%%%%%%%%%%%%%%%%%
%%Okayama U
\name{M.~Harada}{\AFFokayama},
\name{Y.~Koshio}{\AFFokayama,\AFFipmu},
%
%%%%%%%%%%%%%%%%%%%%%%%%%%%%%%%%%%%%%%%%%%%%%%%%%%%%%%%%%%%%%%%%%%%%
%%Sheffield
\name{M.~D.~Thiesse}{\AFFsheff},
\name{L.~F.~Thompson}{\AFFsheff},
\name{P.~R.~Scovell}{\AFFbul},
\name{E.~Meehan}{\AFFbul},
%%%%%%%%%%%%%%%%%%%%%%%%%%%%%%%%%%%%%%%%%%%%%%%%%%%%%%%%%%%%%%%%%%%%%
%%Tohoku, 
\name{K.~Ichimura}{\AFFtohoku},
\name{Y.~Kishimoto}{\AFFtohoku},
%%%%%%%%%%%%%%%%%%%%%%%%%%%%%%%%%%%%%%%%%%%%%%%%%%%%%%%%%%%%%%%%%%%%%
%%Tokyo, Deartment of Physics
\name{Y.~Nakajima}{\AFFtodai},
%%%%%%%%%%%%%%%%%%%%%%%%%%%%%%%%%%%%%%%%%%%%%%%%%%%%%%%%%%%%%%%%%%%%%
%%IPMU
\name{M.~R.~Vagins}{\AFFipmu,\AFFuci},
%%%%%%%%%%%%%%%%%%%%%%%%%%%%%%%%%%%%%%%%%%%%%%%%%%%%%%%%%%%%%%%%%%%%%
%%TUS
\name{H.~Ito}{\AFFtus},
%%Tukuba 
\name{Y.~Takaku}{\AFFtsukuba},
%%%%%%%%%%%%%%%%%%%%%%%%%%%%%%%%%%%%%%%%%%%%%%%%%%%%%%%%%%%%%%%
%%%%%%
%%%%%%%%%%%%%%%%%%%%%%%%%%%%%%%%%%%%%%%%%%%%%%%%%%%%%%%%%%%%%%%%%%%%%
%%NYC
\name{Y.~Tanaka}{\AFFnyc},
\name{Y.~Yamaguchi}{\AFFnyc}
}
%%%%%%%%%%%%%%%%%%%%%%%%%%%%%%%%%%%%%%%%%%%%%%%%%%%%%%%%%%%%%%%%%%%%%
%\address{
\affil[\AFFicrr]{{Kamioka Observatory, Institute for Cosmic Ray Research, University of Tokyo, Kamioka, Gifu 506-1205, Japan}}
\affil[\AFFmad]{{Department of Theoretical Physics, University Autonoma Madrid, Madrid, 28049, Spain}}
\affil[\AFFlsc]{{Laboratorio Subterraneo de Canfranc, Canfranc Estacion, Huesca, 22880, Spain}}
\affil[\AFFkek]{{High Energy Accelerator Research Organization (KEK), Tsukuba, Ibaraki 305-0801, Japan }}
\affil[\AFFokayama]{{Department of Physics, Okayama University, Okayama, Okayama 700-8530, Japan }}
\affil[\AFFsheff]{{Department of Physics and Astronomy, The University of Sheffield, Sheffield, S3 7RH, United Kingdom}}
\affil[\AFFbul]{{Boulby Underground Laboratory, Saltburn-by-the-Sea, Redcar \& Cleveland, TS13 4UZ, United Kingdom}} 
\affil[\AFFtohoku]{{Tohoku University Research Center for Neutrino Science, Sendai, Miyagi 980-8578, Japan}}
\affil[\AFFtodai]{{Department of Physics, University of Tokyo, Bunkyo, Tokyo 113-0033, Japan }}
\affil[\AFFipmu]{{Kavli Institute for the Physics and Mathematics of the Universe (WPI), The University of Tokyo Institutes for Advanced Study, University of Tokyo, Kashiwa, Chiba 277-8583, Japan }}
\affil[\AFFuci]{{Department of Physics and Astronomy, University of California, Irvine, Irvine, CA 92697-4575, USA }}
\affil[\AFFtus]{{Department of Physics, Faculty of Science and Technology, Tokyo University of Science, Noda, Chiba 278-8510, Japan }}
\affil[\AFFtsukuba]{{Faculty of Pure and Applied Science, University of Tsukuba, Tsukuba, Ibaraki 305-8577, Japan}}
\affil[\AFFnyc]{{Nippon Yttrium Co., Ltd., Omuta, Fukuoka 836-0003, Japan}}

%\author{Insert second author name here}
%\affil{Insert second author address here}

%\author{Insert third author name here}
%\author[3]{Insert fourth author name here} %%% Use optional bracket [3] to change the respective address
%\affil{Insert third author address here}
% ミーティングメンバー＋NYC
%　boulby, canfranc も入れる

%\author{Insert last author name here\thanks{These authors contributed equally to this work}}
%\affil{Insert last author address here}

% screening paper, Okada, NYC, Pablo, Mat.,  screeninig member SIto,HIto

%%% To include the collaborator name... Please use the command "\collaborator"
%%% For example: \collaborator{ATLAS Collaboration}

\begin{abstract}%
This paper reports the development and detailed properties of about 13 tons of gadolinium sulfate octahydrate, \GdSOw, which has been dissolved into Super-Kamiokande (SK) in the summer of 2020. We evaluate the impact of radioactive impurities in \GdSOw\ on DSNB searches and solar neutrino observation and confirm the need to reduce radioactive and fluorescent impurities by about three orders of magnitude from commercially available high-purity \GdSOw. In order to produce  ultra-high-purity \GdSOw, we have developed a method to remove impurities from gadolinium oxide, Gd$_2$O$_3$, consisting of acid dissolution, solvent extraction, and pH control processes, followed by a  high-purity sulfation process. All of the produced ultra-high-purity \GdSOw\ is assayed by ICP-MS and HPGe detectors to evaluate its quality. Because of the long measurement time of HPGe detectors, we have employed several underground laboratories for making parallel measurements including LSC in Spain, Boulby in the UK, and Kamioka in Japan. In the first half of production, the measured batch purities were found to be consistent with the specifications. However, in the latter half, the \GdSOw\ contained one order of magnitude more $^{228}$Ra than the budgeted mean contamination. This was correlated with the corresponding characteristics of the raw material Gd$_2$O$_3$, in which an intrinsically large contamination was present. Based on their modest impact on SK physics, they were nevertheless introduced into the detector. To reduce $^{228}$Ra for the next stage of Gd loading to SK, a new process has been successfully established.

\end{abstract}

\subjectindex{xxxx, xxx}

\maketitle

%% Target  drafting    June  half complete   July complete
%% Target submission    Sep

\section{Introduction} %%Ikeda/Nakajima

%The Super-Kamiokande (SK) detector was upgraded by dissolving Gadolinium (Gd) into its pure water and started an entirely new phase of observation as SK-Gd. 
Super-Kamiokande (SK)~\cite{Super-Kamiokande:2002weg} is a 50-kton water Cherenkov detector located in Hida-city, Japan and has been operational since 1996. The experiment started with pure water as the detection medium and has made world-leading measurements of astrophysical, atmospheric and accelerator neutrinos, as well as searches for nucleon decays. To further enhance its physics capability, such as searching for the Diffuse Supernova Neutrino Background (DSNB), it was proposed to add gadolinium to the SK detector (SK-Gd)~\cite{Beacom:2003nk}.
Gadolinium, in the form of \GdSOw, is expected to significantly enhance the neutron detection capability of SK thanks to its large thermal neutron capture cross section and subsequent $\sim$8~MeV gamma-ray emission.

While one of the primary physics goals of SK-Gd is to detect the DSNB signal via inverse-beta reactions ($\bar{\nu}_{e}+ p \to e^+ + n$), measurements of conventional physics targets will continue
%, such as solar neutrinos, atmospheric neutrinos, beam neutrinos and nucleon decays, 
with similar sensitivities as before. 
One challenge to overcome in the realization of the SK-Gd project is to mass-produce many tonnes of \GdSOw\ with impurities low enough so that, when dissolved into SK, backgrounds for the solar neutrino analysis at SK would not increase significantly and the detector performance would not be otherwise adversely affected.
To achieve this, we developed a scalable purification method and conducted highly-sensitive radiopurity assays at multiple sites.
These efforts were critical for the successful first Gadolinium loading to the SK detector in 2020~\cite{Super-Kamiokande:2021the}.

In this paper, we first present and justify the requirements imposed on radioactive impurities in the \GdSOw\ in Section~\ref{requirements}. Section~\ref{market_production} describes the level of typical impurities of commercially available \GdSOw, which motivated significant R\&D of the purification methods described in Section~\ref{sec:purification}. We describe the low-background, low-activity radioimpurity assay methods in Section~\ref{sec:screening} and their results in Section~\ref{sec:results}.
Finally, we describe further improved purification methods, which will be employed for future Gadolinium loading in Section~\ref{sec:new_purification}.

%\section{Requirements}\label{requirements}  %% Pablo 
\section{Requirements}\label{requirements}

While the addition of Gd to SK will allow more efficient neutron tagging, any impurities present in the raw \GdSOw\ would also be present in the detection medium. Radioactive impurities may contribute to the low energy backgrounds and affect the sensitivity of SK-Gd to the observation of the DSNB and solar neutrino measurements (4-20 MeV). To maintain these sensitivities, stringent requirements are set on the radioactive purity of the \GdSOw\ material to be introduced to SK.

In \ref{sec:usf}, we consider gamma emission accompanied by a neutron from radioactive contaminants; these neutrons are produced throughout the detector and their detection could mimic an antineutrino inverse beta decay signal.

Further, in \ref{sec:betas} and \ref{sec:nrad}, the relevant radioactive impurities to be controlled include those which emit gamma or beta radiation with sufficient energy to mimic signals from low energy, $\mathcal{O}$(MeV), neutrino interactions.

In this section, we study the radioactivity-induced backgrounds for SK-Gd and set the requirements on the \GdSOw\ impurities based on the physics analyses they impact. We also justify the limit imposed on the fluorescent cerium concentration to prevent a significant adverse change in detector performance. 

For all these studies we assume 0.2\% concentration of \GdSOw, or 0.1\% concentration of Gd.

\subsection{$^{238}$U Spontaneous Fission}\label{sec:usf}
A small fraction of $^{238}$U decays are the result of spontaneous fission (SF), as are even smaller fractions of $^{235}$U and $^{232}$Th due to their lower masses. A neutron captured on Gd and the resulting gamma cascade following an SF event can be indistinguishable from an antineutrino inverse-beta decay (IBD) if only one gamma and one neutron are detected. Therefore, these decays represent an irreducible source of background for measuring low-energy antineutrinos from the DSNB, nuclear reactors, and the silicon-burning phase prior to a supernova burst. 

The number of $^{238}$U decays which fake an antineutrino IBD is calculated using several factors. First, the likelihood of $^{238}$U to decay by SF is $5.4\cdot10^{-7}$~\cite{popeko80}. Second, the target concentration of \GdSOw\ in SK-Gd is 0.2\%. And third, the number of high energy, $\mathcal{O}(\text{MeV})$, photons produced with energy $E_{\rm min} \leq E_\gamma \leq E_{\rm max}$ in an SF decay, as derived from the emission spectrum measured in \cite{Sobel73}, is:
\begin{equation}\label{eq:sfphot}
    N^\gamma_{SF}\left(E_{\rm min},E_{\rm max}\right)=0.99\cdot \left( e^{-\frac{E_{\rm min}}{1.41}}-e^{-\frac{E_{\rm max}}{1.41}}\right),
\end{equation}
where the factors $0.99$ and $1.41$ are the normalization and slope of the exponential fit to the data in~\cite{popeko80} of gamma emission rate from $^{238}$U SF as a function of the energy. One should also note that for the photons to be detected as electrons in SK, they must have 0.511~MeV additional energy to account for the electron mass.

For an SF decay to mimic an antineutrino IBD, one emitted neutron must be detected. Combining the neutron multiplicity probability~\cite{popeko75,Ethvignot05} from $^{238}$U SF with the SK estimated neutron tagging efficiency of 80\% (given the 0.2\% concentration of \GdSOw) along with calculations in \cite{FernandezMenendez:2017ccn}, the irreducible background rate (events per second) for antineutrino IBD events from $^{238}$U SF is
\begin{equation}\label{eq:usf}
R_{^{238}U-SF}\left(1\gamma+1\text{n}\right) = \left(5.4\cdot10^{-7}\right) \cdot 0.364 \cdot \mathcal{C}_{^{238}\text{U}} \cdot 0.2\% \cdot m_{\rm FV} \cdot N^\gamma_{SF}\left(E_{\rm min},E_{\rm max}\right), 
\end{equation}
where $5.4\cdot10^{-7}$ is the $^{238}$U SF rate per decay, $m_{\rm FV}$ is the mass of the SK fiducial volume (around $22.5 \cdot 10^6$ kg depending on the physics analysis),
% 0.002 is the concentration of \GdSOw\ in the SK water (0.2\%), 
$\mathcal{C}_{^{238}U}$ is the specific activity of $^{238}$U in the \GdSOw\ in units of Bq/kg and 0.2\% is the target \GdSOw\ concentration in SK-Gd.
% $\rm BR_{^{238}U_{\rm SF}}$ is the fraction of SF of $^{238}U$ decays 
The factor $0.364$ is the fraction of $^{238}$U SF decays with a single reconstructed neutron. This accounts for the neutron multiplicity distribution measured in~\cite{popeko75} and the above 80\% neutron tagging efficiency, which depends on the target \GdSOw\ concentration.

For the SK DSNB analysis, events with reconstructed energy from $10-20$~MeV are considered. For reactor neutrinos, the photon energies relevant are from $3-10$~MeV, where the minimum energy is limited by the SK low energy threshold.

The irreducible backgrounds from $^{238}$U SF which mimic antineutrino IBD events impact the reactor antineutrino detection analysis, the DSNB measurement, and the detection of pre-supernova antineutrinos at SK. Among these analyses, the detection of the DSNB is most constrained by backgrounds from $^{238}$U SF~\cite{FernandezMenendez:2017ccn}. 
%Therefore, the radiopurity requirements for the early part of the $^{238}$U decay chain in \GdSOw\ are motivated by this source of background.

Depending on the theoretical model for the DSNB, the expected signal rate at SK ranges from 0.6 to 5.3 events per year with antineutrino energy above 10~MeV \cite{PhysRevD.104.122002}. To maintain a favourable signal-to-background ratio, the concentration of $^{238}$U in dissolved \GdSOw\ is limited to less than 5~mBq/kg, which corresponds to about 0.8 events per year in the SK fiducial volume from Equation~\ref{eq:usf}.

\subsection{$\beta$-rays from Radioactivity}\label{sec:betas}
Sources of $\beta$ radiation in SK primarily impact the solar neutrino analysis because those $\beta$s mimic low-energy electron neutrino interactions. Additionally, a $\beta$ decay when coincident with a neutrino interaction may mimic an antineutrino IBD event. The most significant $\beta$ decay backgrounds are those that come from decays with a large branching fraction in the respective decay chain and a high $Q$-value: $^{208}$Tl ($Q_\beta = 5.00$ MeV, 35.9\% of $^{228}$Th decays), $^{212}$Bi ($Q_\beta = 2.25$ MeV, 64.1\% of $^{228}$Th decays) and $^{214}$Bi ($Q_\beta = 3.27$ MeV, nearly 100\% of $^{226}$Ra decays). These $\beta$s may be reconstructed as the signal of low energy neutrinos for the lowest energy bins in the SK solar neutrino analysis. According to \cite{FernandezMenendez:2017ccn}, the fraction of these $\beta$ decays which pass the SK solar neutrino cuts are 0.19\% for $^{208}$Tl, 0.01\% for $^{214}$Bi and $<2\cdot 10^{-4}$\% for $^{212}$Bi.

The expected solar neutrino candidate event rate in SK is around 200 per day with recoil electron kinetic energy between 3.5~MeV and 10~MeV~\cite{PhysRevD.94.052010}. From the above estimates, 0.05~mBq/kg of $^{208}$Tl contamination will produce about 130 solar neutrino candidate background events per day. About 190 candidate background events is expected from the $\beta$ decay of  $^{214}$Bi in the late part of the $^{238}$U chain ($^{226}$Ra equilibrium) for a 0.5~mBq/kg contamination. 

\subsection{Radioactivity-Induced Neutron Production}\label{sec:nrad}
$^{238}$U SF is the main neutron producer in the SK fiducial volume. For these decays, 67\% of the final states contain more than one neutron~\cite{popeko75}, making these background events easily distinguishable from solar neutrino candidates considering the high neutron tagging efficiency in SK-Gd.

However, $(\alpha,n)$ reactions on oxygen from all $\alpha$ decays in all radioactive chains comprise another irreducible source of background for solar neutrino candidate events in SK as the emitted neutrons are captured by Gd, producing an 8~MeV $\gamma$ cascade. In fact, 40\% of neutron captures on Gd pass the SK solar neutrino cuts~\cite{FernandezMenendez:2017ccn}.

As $^{18}$O is naturally five times more abundant than $^{17}$O, these reactions occur mostly on $^{18}$O nuclei. In these reactions, neutrons are produced in pairs, 
\begin{equation}
    ^{18}\rm O+^4\alpha\longrightarrow ~^{22}Ne^*\longrightarrow ~^{20}Ne+2n,
\end{equation}
which together with the high neutron capture efficiency of Gd, helps reduce this source of background for solar neutrinos.

\begin{figure*}
     \centering
     \begin{minipage}[t]{0.48\textwidth}
         \centering
         \includegraphics[width=\textwidth]{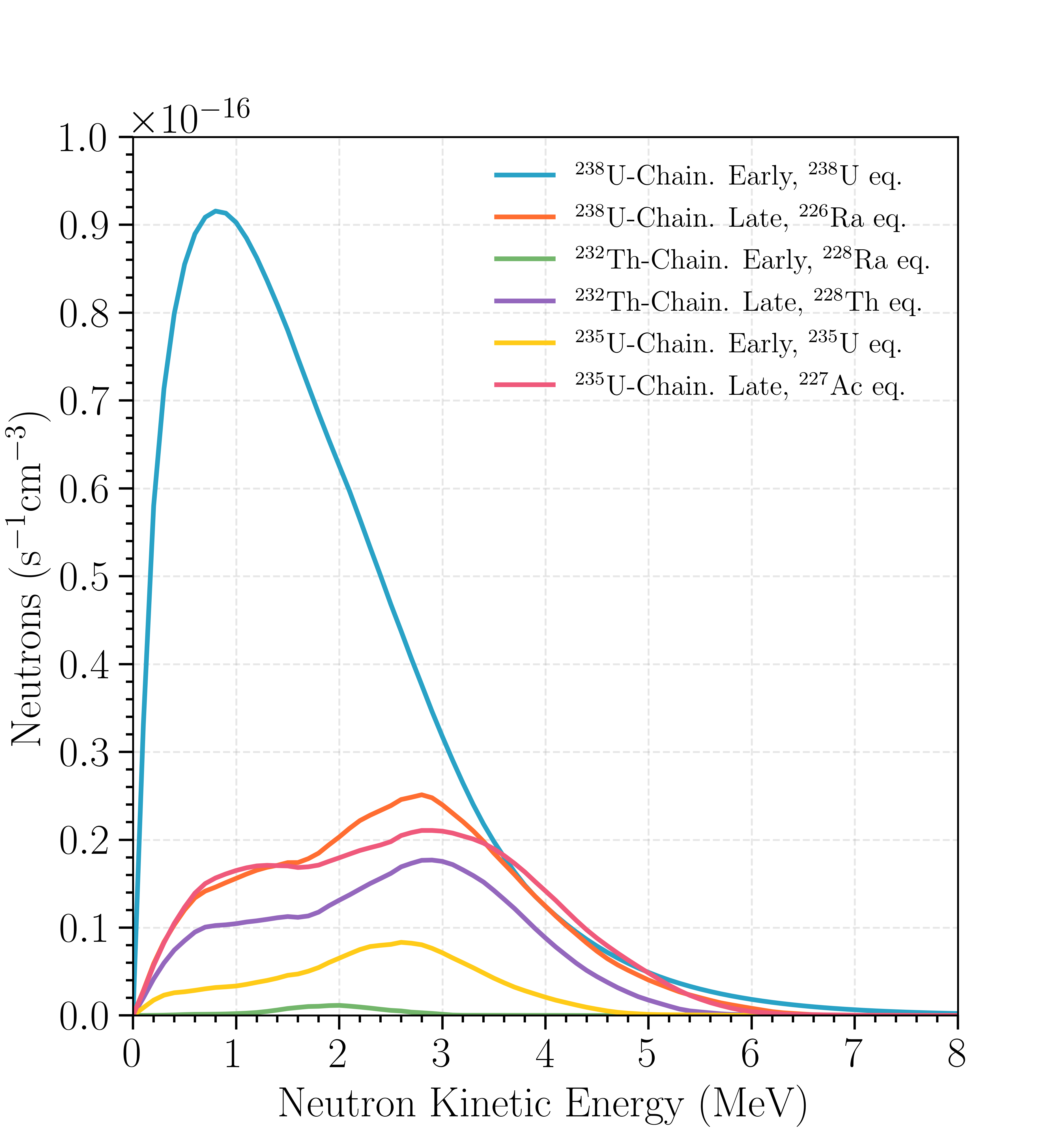}
         \captionof{figure}{\label{fig:neutronspec_bynuclide} The neutron production spectrum from each part, Early and Late, of the $^{238}$U, $^{232}$Th and $^{235}$U chains assuming secular equilibrium, an activity of 1~mBq/kg each and 0.2\% \GdSOw.}
     \end{minipage}
     \hfill
     \begin{minipage}[t]{0.48\textwidth}
         \centering
         \includegraphics[width=\textwidth]{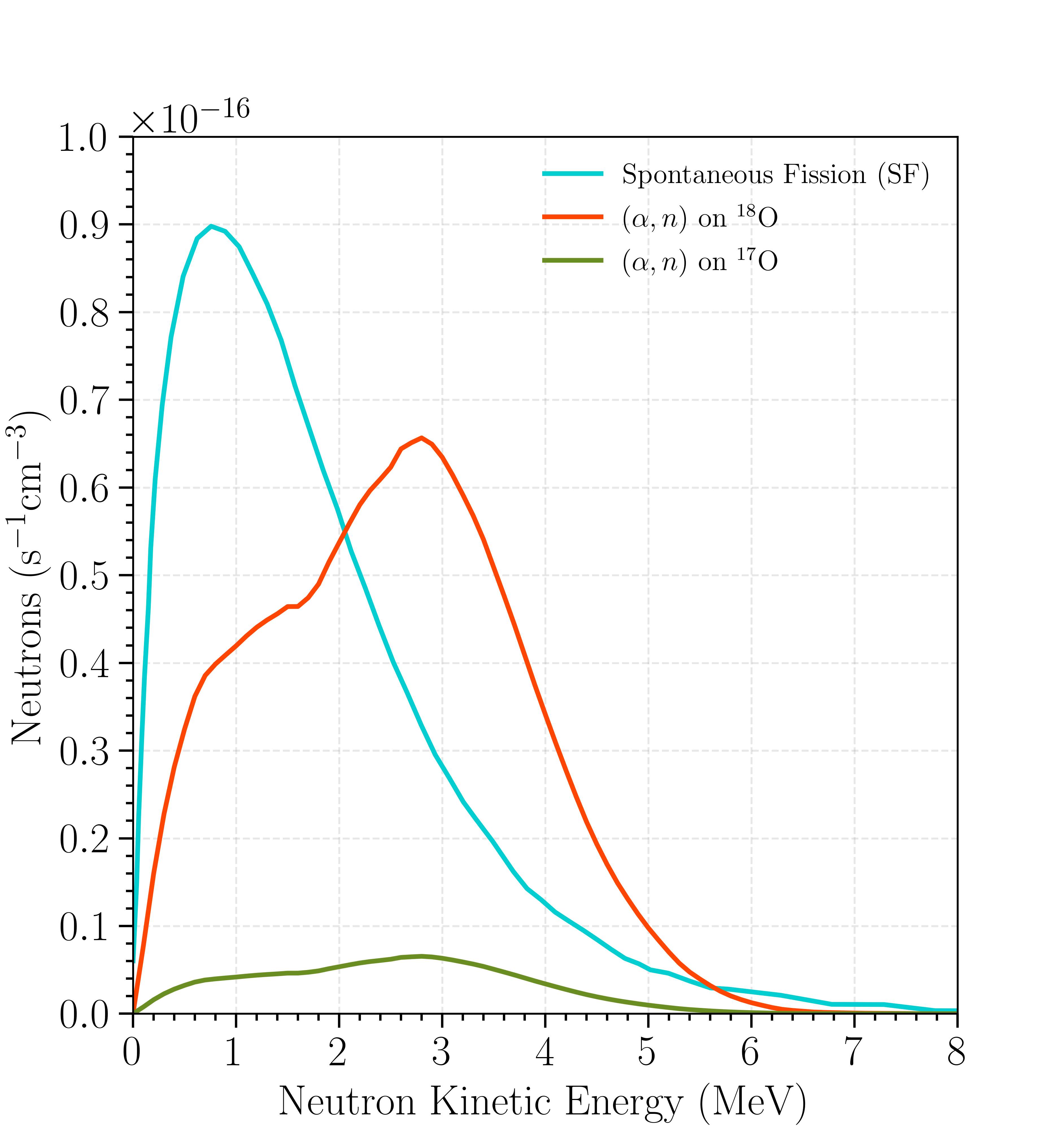}
         \captionof{figure}{\label{fig:neutronspec_byprocess} The neutron production spectrum from each contributing process given 1~mBq/kg of each $^{238}$U, $^{232}$Th and $^{235}$U at a concentration of 0.2\% \GdSOw.}
    \end{minipage}
%\caption{\label{fig:neutronspec} }
\end{figure*}

To quantify the production of neutrons from SF and $(\alpha,n)$ reactions, we use the SOURCES-4C code~\cite{wilson02}. The neutron production rates for an activity of 1~mBq/kg of each impurity decay chain and 0.2\% \GdSOw\ concentration are shown in Figure~\ref{fig:neutronspec_bynuclide} for each decay sub-chain and in Figure~\ref{fig:neutronspec_byprocess} for every relevant reaction. The total expected neutron production rate (events per second) from $\alpha$ decays is then parameterized by:
\begin{equation}\label{eq:radneutrons}
\begin{split}
R_{neutrons}^{rad} (1\text{n})=& \left(2.31\cdot 10^{-9}\cdot\mathcal{C}_{^{238}\rm U_{early}} + 0.80\cdot 10^{-9}\cdot\mathcal{C}_{^{238}\rm U_{late}}\right.\\
&+ 0.02\cdot 10^{-9}\cdot\mathcal{C}_{^{232}\rm Th_{early}} + 0.54\cdot 10^{-9}\cdot\mathcal{C}_{^{232}\rm Th_{late}}\\ 
&\left.+ 0.20\cdot 10^{-9}\cdot\mathcal{C}_{^{235}\rm U_{early}} + 0.78\cdot 10^{-9}\cdot\mathcal{C}_{^{235}\rm U_{late}}\right)   \cdot 0.461 \cdot
m_{\rm FV},
\end{split}
\end{equation}
where the prefactors on each term (in units of neutrons per second per kg of water per Bq/kg of contamination) represent the total neutron production rate for that sub-chain given the 0.2\% \GdSOw\ concentration and integrated over the full neutron kinetic energy range. $\mathcal{C}_x$ corresponds to the concentration of radioactive impurities for each part of each decay chain in units of Bq/kg and the $0.461$ factor is the fraction of $(\alpha,n)$ reactions which result in at least one of the two emitted neutrons being captured on Gd (90\% capture efficiency) but only one reconstructed as a solar neutrino candidate (40\% chance to pass SK solar neutrino event cuts). 

Background neutrons from $(\alpha,n)$ reactions on $^{18}$O impact the solar neutrino analysis at SK. At the impurity concentration imposed on early-chain $^{238}$U by the DSNB analysis and on the $^{226}$Ra and $^{228}$Th chains by $\beta$ emission as backgrounds to the solar neutrino analysis, the radioactivity-induced neutrons from these chains do not impose any stricter requirements on the impurity concentrations.

For early-chain $^{232}$Th, neither SF nor $(\alpha,n)$ reactions on $^{18}$O comprise a significant portion of background events on any physics analysis. However, because the half-lives of $^{228}$Ra and $^{228}$Th are on the same scale as the lifetime of the SK-Gd experiment, we impose the same requirement on the early part of the chain as imposed by $\beta$ emission on the late part of the chain, $<0.05$~mBq/kg. 

Similarly, $^{235}$U is not significantly limited by SF or $\beta$ emission. To derive a limit on $^{235}$U activity we consider that both $\beta$ emission and radioactivity-induced neutron production contribute to the solar neutrino background. Since the background candidate event rate from $\beta$ emission of $^{226}$Ra and $^{228}$Th are together greater than 300 events per day, we limit $^{235}$U to as low as reasonably possible. A limit of $<30$~mBq/kg, which is easily achievable through purification, produces around 26 background candidate events per day. We adopt this as the impurity requirement for the entire $^{235}$U decay chain.

%$<5$~mBq/kg, about 10 candidate background events per day are expected as background to the solar neutrino analysis. Similarly, the candidate background event rates arising from $(\alpha,n)$ reactions from the $^{226}$Ra and $^{228}$Th chains are several orders of magnitude lower than their contributions to the solar neutrino analysis background from their $\beta$ emission. The impurity concentration of early-chain $^{232}$Th is not significantly limited by SF or $\beta$ emission.

%As is the case for high energy $\beta$s, background neutrons from $(\alpha,n)$ reactions on $^{18}$O impact the solar neutrino analysis at SK. However the relative impact of each background type and source nuclide on each analysis is different. For instance, the candidate background event rates arising from $(\alpha,n)$ reactions from the $^{226}$Ra and $^{228}$Th chains are several orders of magnitude lower than their contributions to the solar neutrino analysis background from their $\beta$ emission. For $^{235}$U, however, 30~mBq/kg of $^{235}$U produces 26 solar neutrino candidate neutron background events per day from Equation~\ref{eq:radneutrons} and does not significantly contribute to the DSNB or solar neutrino analysis backgrounds in other ways. Therefore, we set 30~mBq/kg as the limit for $^{235}$U contamination.

\subsection{Fluorescence of Ce}

The Cherenkov light wavelength shifting properties of cerium ions have a negative impact on the response of the SK detector. Cerium is typically found in \GdSOw\ at concentrations at the parts-per-million (ppm) level. The excitation maximum for cerium ions is 255~nm and the fluorescence emission maximum is around 350~nm with a time constant of around 100~ns~\cite{shigematsu1971}. Cherenkov light is emitted in a continuous spectrum that is inversely proportional to the square of the wavelength, and the direction of the photons correlates with the direction of the incoming particles. However, when Cherenkov light near 255~nm, to which the SK photosensors are insensitive, is re-emitted and observed in a random direction with a time constant of 100~ns due to wavelength conversion of Ce, the pattern and time resolution of the longer wavelength Cherenkov light degrades. Therefore, it is necessary to reduce the cerium concentration in SK-Gd as much as possible. 

It was found that the commercially available \GdSOw\ used in our R\&D with the EGADS detector~\cite{Marti:2019dof} contained about 50~ppm Ce. The observed time distribution of Cherenkov light was distorted significantly by the light emission by Ce. It has been confirmed by EGADS that the hit timing disruption of Ce is not observed when the Ce concentration of \GdSOw\ is reduced by three orders of magnitude from the nominal value. Therefore, the requirement of Ce content in \GdSOw\ for SK-Gd was set to 50 parts-per-billion (ppb).

Although Gd and Ce are easily separated to the ppm level in the process of refining rare earths, a special treatment is added for a further reduction of Ce to the ppb level or below, which will be explained in Section~\ref{sec:Neutralization_and_sulfation}. 

\subsection{Summary of Requirements}

SK-Gd aims to measure low energy neutrinos from the Sun and antineutrinos from the DSNB, nuclear reactors, and the silicon-burning stage prior to a supernova burst. Among them, the solar neutrino and DSNB antineutrino measurements set the strongest requirements on the purity of the \GdSOw.

The detection of the DSNB is most strongly limited by $^{238}$U SF mimicking an antineutrino IBD signal. Thus, the concentration limit imposed on the early part of the $^{238}$U decay chain in \GdSOw\ is $<5$~mBq/kg.

Because the ranges of the lowest-energy detectable solar neutrinos and the highest-energy $\beta$ rays overlap, the detection of solar neutrinos drives the allowable concentration of the $^{232}$Th chain, $<0.05$~mBq/kg, and the late part of the $^{238}$U chain, $<0.5$~mBq/kg. For $^{235}$U, the detection of solar neutrinos is most strongly limited by radioactivity-induced neutron production, thereby limiting this nuclide to $<30$~mBq/kg.

The radioactivity contamination requirements for \GdSOw\ are summarized in Table~\ref{table:requirements}.

\begin{table}[ht]
\begin{center}
\begin{tabular}{c|c|c|c}
\hline
Chain & Part of chain & Requirement (mBq/kg) & Requirement (ppb) \footnotemark{} \\
\hline\hline
\multirow{2}*{$^{238}$U} & $^{238}$U & $<~5$ & $<~0.4$\\ 
& $^{226}$Ra & $<~0.5$  &  \\ 
\hline
\multirow{2}*{$^{232}$Th} & $^{232}$Th & $<~0.05$ & $<~0.013$\\ 
& $^{228}$Th & $<~0.05$ & \\ 
\hline
\multirow{2}*{$^{235}$U} & $^{235}$U & $<~30$ & $<~50$\\ 
& $^{227}$Ac / $^{227}$Th & $<~30$ & \\ 
\hdashline
\multicolumn{2}{c|}{Ce} & -- & $<~50$\\
\hline
\end{tabular}
\caption{Summary of contamination requirements of \GdSOw\ in units of mBq per kilogram and ppb based on the needs to perform DSNB and solar neutrino analyses.}
\label{table:requirements}
\end{center}
\end{table}
\footnotetext{ppb values are in units of kg/kg and corrected assuming each isotope's natural abundance.}

Additionally, once the \GdSOw\ is dissolved into the water, the uranium concentration is expected to be reduced by approximately two orders of magnitude using the resin-based filtration within the water purification system~\cite{Marti:2019dof,ABE2022166248}. This has a further positive impact on the measurements affected by uranium contamination. 

%\section{Market production}\label{market_production} %% Luis
\section{Market production}\label{market_production} %% 

The world market was searched for suppliers of high purity \GdSOw. Several samples were purchased from different companies to measure and compare their radioactive contaminants. These measurements were performed at Canfranc Underground Laboratory (LSC) in Spain (see Section~\ref{lsc}), and the results summarized in Table~\ref{tab:market}.

\begin{landscape}

\begin{table}[htp]
\centering
%\tiny
%\small
\begin{tabular}{lc|cc|cc|cc|ccc}
 & & \multicolumn{2}{c|}{ $^{238}$U Chain } & \multicolumn{2}{c|}{$^{232}$Th  Chain} & \multicolumn{2}{c|}{$^{235}$U Chain} & \multicolumn{3}{c}{Others}  \\ \hline
% & & \multicolumn{1}{l}{E} & \multicolumn{1}{l}{L} & \multicolumn{1}{l}{E} & \multicolumn{1}{l}{L}  & \multicolumn{1}{l}{E}  & \multicolumn{1}{l}{L}  &  &  &  \\
Company &
  Date &  E, $^{238}$U eq. & L, $^{226}$Ra eq. & E, $^{228}$Ra eq. & L, $^{228}$Th eq.  & E, $^{235}$U eq.  & L, $^{227}$Ac eq. & $^{40}$K & $^{138}$La & $^{176}$Lu  \\
  \hline\hline
  A (USA) & 2009/04 & 51$\pm$21  &  8$\pm$1 & 11$\pm$2 & 28$\pm$3 & $<$32  & 214$\pm$10 &  29$\pm$5  & 8$\pm$1  & 80$\pm$8 \\
  A (USA) & 2010/08 & $<$33 & 2.8$\pm$0.6 & 270$\pm$16 & 86$\pm$5 & $<$32& 1700$\pm$20 & 12$\pm$3 & - & 21$\pm$2\\
  B (China) & 2012/08 & 292$\pm$6 & 74$\pm$2 & 1099$\pm$12 & 504$\pm$6 & $<$112 & 2956$\pm$30 & 101$\pm$10 & 683$\pm$15 & 566$\pm$6 \\
  C (China) & 2013/02 & 74$\pm$28 & 13$\pm$1 &  205$\pm$6 & 127$\pm$3 & $<$25 & 1423$\pm$21 & 60$\pm$7 & 3$\pm$1 & 12$\pm$1 \\
  B (China) & 2013/03 & 242$\pm$6 & 13$\pm$2 & 21$\pm$3 &  374$\pm$6 & $<$25 & 175$\pm$42 & 18$\pm$8 & 42$\pm$3 &  8$\pm$2 \\
  A (USA) & 2013/08 & 71$\pm$20 & 8$\pm$1 & 6$\pm$1 & 159$\pm$3 & $<$32 &  295$\pm$10 & 3$\pm$2  & 5$\pm$1 & 30$\pm$1 \\
  D (China) & 2013/07 & 47$\pm$26 & 5$\pm$1 & 14$\pm$2 & 13$\pm$1 & $<$12 &$<$6 & 3$\pm$2 & $<$1 & 1.6$\pm$0.3 \\
  D (China) & 2013/07 &  73$\pm$27& 6$\pm$1 & 3$\pm$1& 411$\pm$5 & $<$30 &  $<$18 &  8$\pm$4 & $<$1  & $<$2 \\
  A (USA) & 2014/02 &  $<$ 76  &  $<$ 1.4 & 2$\pm$1 & 29$\pm$2 & $<$1.8 & 190$\pm$6 &  $<$5 &  23$\pm$1 & 2.5$\pm$0.6  \\
\end{tabular}
\caption{
Radioactivity in samples of \GdSOw\ obtained from different suppliers at different times, as measured at the LSC. Activities presented are in units of mBq/kg and limits are at 95\% C.L. The measurements of the radioactive chains are separated into those for the early part of the chain (E) and the late part of the chain (L). The isotopes quoted are the longest lived within the sub-chain, and the activities are estimated assuming equilibrium (eq.)}
\label{tab:market}
\end{table}

\end{landscape}

These measurements indicate the presence of non-negligible naturally-occurring radioactive contamination in the material, much higher than the requirements discussed in the previous section. It is worth noting that the early and late parts of the radioactive decay chains do not appear to be in secular equilibrium. Table~\ref{tab:market} clearly shows that a dedicated R\&D program must be carried out to purify the available \GdSOw\ powder to levels which satisfy the requirements in Table~\ref{table:requirements}.

The commercially available \GdSOw\ used in our R\&D with the EGADS detector~\cite{Marti:2019dof} contained about 50~ppm Ce, which is near the natural abundance of cerium in the Earth's crust.

\section{Purification method}\label{sec:purification} %%Ikeda

\subsection{Gd-Oxide and its solution by acid}

Gadolinium oxide, \GdOx, is used as a raw material for \GdSOw\ production. The purity of \GdOx\ is better than 99.99\% by mass as \GdOx\ / TREO (gadolinium oxide amount in total rare earth oxide amount). The concentration of trace uranium and thorium contained in \GdOx\ with this purity is typically about 500-1000~ppb, but we carefully selected raw material with a low thorium concentration ($150-200$~ppb) for this development.

To dissolve \GdOx, hydrochloric acid is used. This is because gadolinium chloride, GdCl$_3$, has a high solubility, and wastewater treatment is relatively easy since it does not contain nitrogen. This gadolinium acidic aqueous solution becomes a raw material for the subsequent production process after the insoluble matter is filtered off. The acid concentration is set to 30-40\% by weight, resulting in high reactivity and low volatility. From the viewpoint of the removal efficiency of radioactive impurities and the yield in the subsequent processes, the target gadolinium concentration in this solution is then $\sim$305 g/L (Gd$_2$O$_3$: 350g/L). 

\subsection{Solvent extraction}

The solution obtained above is then processed with the solvent extraction to increase the purity. While maintaining Gd in the aqueous phase, radioactive impurities such as Th and U can be extracted into the organic solvent phase using extractant 2-ethylhexyl 2-ethylhexylphosphonate (PC-88A manufactured by DAIHACHI Chemical Industry Co., Ltd.). It is known that rare earth elements are extracted and separated in an organic solvent containing 2-ethylhexylphosphonate \cite{KUMARI20181029}. The extraction solvent is diluted with an isoparaffin (IP3835 manufactured by Idemitsu Kosan Co., Ltd.) which reduces the viscosity and specific gravity of the organic solvent phase. Reducing the viscosity improves contact between the aqueous phase and the organic solvent phase to improve the efficiency of solvent extraction. Reducing the specific gravity accelerates the separation between the organic solvent phase and the aqueous phase after mixing. The extraction solvent and the isoparaffin solvent are mixed at a ratio of 20:80.

Between a pH of 1.0 and 1.3, Gd is not extracted into the organic solvent phase, but Th and U are. During this process, the pH is controlled at 1.0 by using an ammonia solution to prevent metal contamination and enhance separation efficiency.  After mixing the isoparaffin with the aqueous solution, the aqueous phase is separated as a purified Gd solution by allowing the mixed solution to separate. This solvent extraction step is performed twice for this production. 

\subsection{Neutralization and sulfation}
\label{sec:Neutralization_and_sulfation}
After solvent extractions, two neutralization processes are performed to reduce Th and Ce.
Because the precipitation pH of thorium hydroxide is lower than the precipitation pH of gadolinium hydroxide, thorium hydroxide can be removed by coprecipitating with a small amount of gadolinium hydroxide by adjusting pH. The pH is adjusted to 4.8 by adding ammonia water to the Gd solution after the solvent extraction.
In an acidic solution, Ce is trivalent and stable but it was precipitated as cerium oxide by adding hydrogen peroxide solution and raising the electric potential in the solution to oxidize it.
By performing this treatment together with the precipitation of thorium hydroxide described above, cerium oxide is coprecipitated together with thorium hydroxide. Hydrogen peroxide is added so that it is 0.3\% in the aqueous solution.

Finally, 98\% purity sulfuric acid is added to the obtained Gd solution to precipitate Gd sulfate. Gd sulfate is filtered to remove the filtrate, and washed with pure water until the pH reaches 4. The produced gadolinium sulfate is octahydrate. This can be confirmed by X-ray diffraction (Appendix~\ref{appendix:xraydiff}) or thermogravimetric analysis (Appendix~\ref{appendix:thermograv}).

Each batch of the \GdSOw\ production is about 500~kg.
The \GdSOw\ is in a powder form but contains an average of 2.5\% additional water left over from processing. So the 13.2~tons of powder dissolved into SK corresponds to 12.9~tons of \GdSOw.

\section{Material Assay} \label{sec:screening}

Samples from each 500~kg batch of \GdSOw\ are assayed to evaluate whether the requirements set out in Section~\ref{requirements} for radioactive and fluorescent impurities are met. Since the nuclides which contaminate \GdSOw\ are not necessarily in secular equilibrium with their long-lived parents and daughters, high purity Germanium (HPGe) gamma spectrometry is used to investigate the activity of the early and late parts of all decay chains which could affect SK-Gd physics sensitivities. Also, inductively-coupled plasma mass spectrometry (ICP-MS) is used to measure the concentration of U, Th, and Ce to very low levels.

ICP-MS is sensitive to the most abundant, long-lived members of the early parts of the U and Th decay chains to the level required by SK-Gd: $^{238}$U to $<5$~mBq/kg and $^{232}$Th to $<0.05$~mBq/kg. HPGe gamma spectrometry infers the activity of long-lived parent nuclides based on the gamma emission of the parent or its daughters. For late-chain $^{238}$U ($^{226}$Ra equilibrium) and all parts of the $^{235}$U chain, only HPGe gamma spectrometry is sensitive to the level specified by the SK-Gd requirements: $<0.5$~mBq/kg and $<30$~mBq/kg, respectively. Neither method of assay used here is sensitive to late-chain $^{232}$Th ($^{228}$Th equilibrium) concentration to $<0.05$~mBq/kg.

\subsection{HPGe Laboratories}

\subsubsection{Boulby}

\begin{figure}[!htb]
\begin{subfigure}{0.48\textwidth}
\centering
\includegraphics[width=\textwidth]{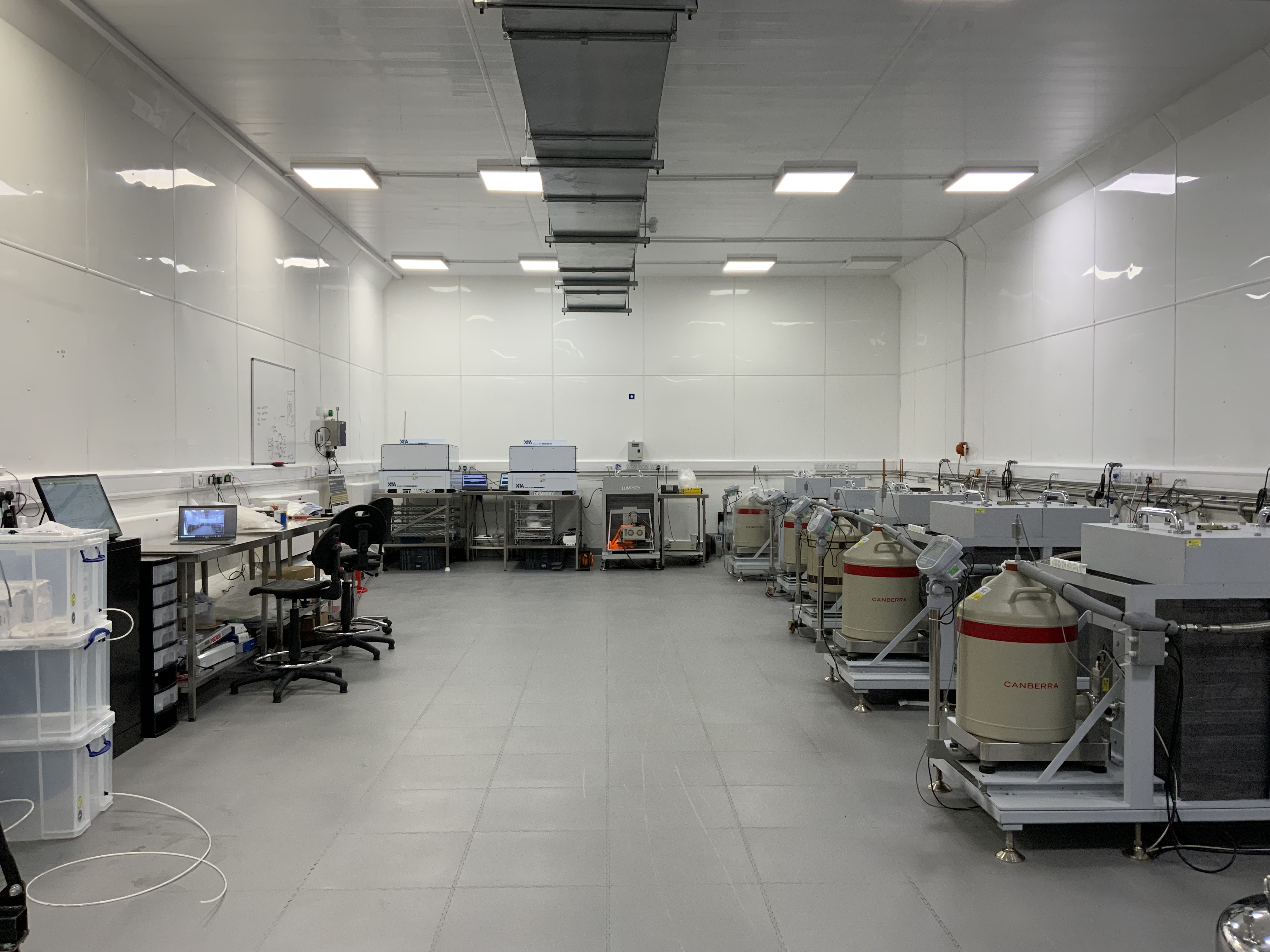}
\caption{\label{fig:bugs}}
\end{subfigure}
\hfill
\begin{subfigure}{0.48\textwidth}
\centering
\includegraphics[width=\textwidth]{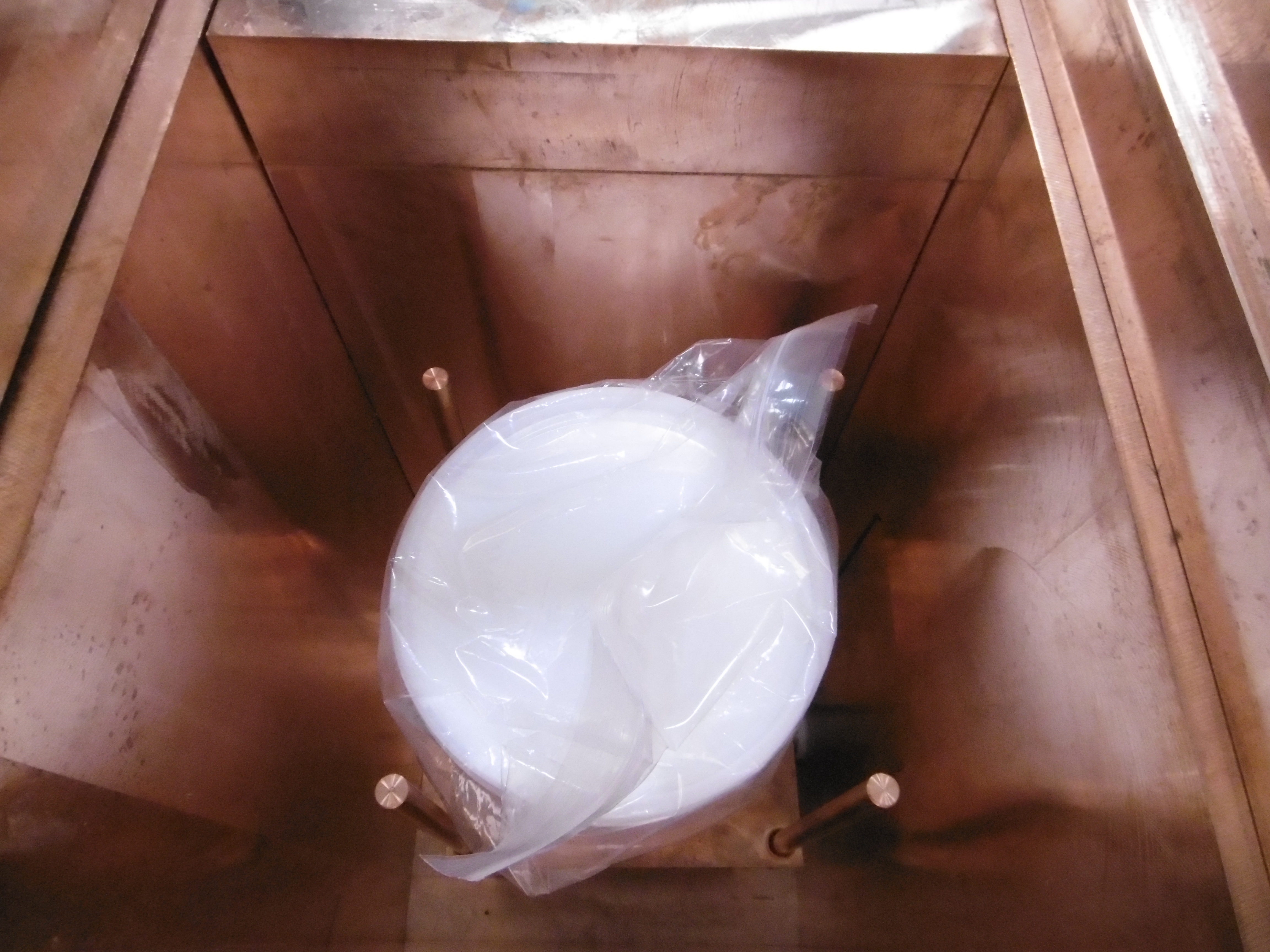}
\caption{\label{fig:merrybent}}
\end{subfigure}
\caption{(a) The BUGS Facility at Boulby Underground Laboratory and (b) a 5~kg \GdSOw\ sample on Merrybent detector.}
\end{figure}

The Boulby UnderGround Screening (BUGS) Facility \cite{scovell2018}  (Fig.~\ref{fig:bugs}) is located 1.1~km underground (2840~m water equivalent) at the Boulby Underground Laboratory in the north of England. Gamma spectrometry of 13 samples of \GdSOw\ was performed using two HPGe detectors at BUGS. Produced by Mirion, the two p-type detectors called Belmont and Merrybent have relative efficiencies of 160\% and 100\%, respectively. The ``specialty ultra-low background'' detectors are housed in similar low-background lead and inner copper-lined shields and are purged with generated $N_2$ gas. Additional details of these detectors and their general performance around the same time as this assay program is reported in \cite{akerib2020lux}.

Samples of 5~kg of \GdSOw\ were packed in Marinelli beakers of type 448G-E from Ga-Ma and Associates, Inc. in a clean environment, then transported into the BUGS class 1000 clean room using standard triple-bagged procedures. Figure~\ref{fig:merrybent} shows a \GdSOw\ sample placed on the Merrybent detector. Samples are measured for several weeks to reach a minimum detectable activity (MDA) of $<0.5$~mBq/kg of $^{226}$Ra-equivalent activity at the 95\% confidence level. 

To achieve the best possible results, several known systematic errors are accounted for in the spectrum analysis using a Geant4-based~\cite{AGOSTINELLI2003250} detector simulation developed in-house. Apart from estimating the gamma detection efficiency for an arbitrary sample geometry, corrections for true coincidence summing~\cite{scovell2018} and background shielding by the sample material~\cite{thiesse2022} are implemented to improve the accuracy of results.

Since this assay program has been completed, several improvements have been made at BUGS to reduce the radon concentration of the nitrogen purge line, greatly improving the sensitivity of Merrybent and Belmont to $^{226}$Ra and its daughters.

%Given the sample size and geometry and the detector backgrounds, the sensitivity of each detector to reach the required SK-Gd detection limits for U and Th is shown in Figure \ref{fig:LD_Boulby}. The lower sensitivity of Merrybent to $^{226}$Ra daughter gammas is caused by a combination of lower detection efficiency and comparatively higher backgrounds than Belmont. 

%\begin{figure}
%\begin{subfigure}{0.375\textwidth}
%\centering
%\includegraphics[width=\textwidth]{ptephy_22_01_16/img/LD_All_Belmont_plot.png}
%\caption{\label{fig:LD_Belmont}}
%\end{subfigure}
%\hfill
%\begin{subfigure}{0.375\textwidth}
%\centering
%\includegraphics[width=\textwidth]{ptephy_22_01_16/img/LD_All_Merrybent_plot.png}
%\caption{\label{fig:LD_Merrybent}}
%\end{subfigure}
%\hfill
%\begin{subfigure}{0.2\textwidth}
%\includegraphics[width=\textwidth]{ptephy_22_01_16/img/LD_legend.png}
%\end{subfigure}
%\caption{$L_D$ as a function of measurement time for (a) Belmont and (b) Merrybent, normalised to the corresponding SK-Gd radioimpurity requirement.}
%\label{fig:LD_Boulby}
%\end{figure}

\subsubsection{Canfranc}
\label{lsc}

The {\it Laboratorio Subterr\'aneo de Canfranc} (LSC) is located on the Spanish side of the Pyrenees Mountains, under the Tobazo peak. The laboratory has a rock shielding of 800~ water equivalent), that suppresses the cosmic muon flux by almost 5 orders of magnitude.

\begin{figure}[!htb]
\begin{subfigure}{0.422\textwidth}
\centering
\includegraphics[width=\textwidth]{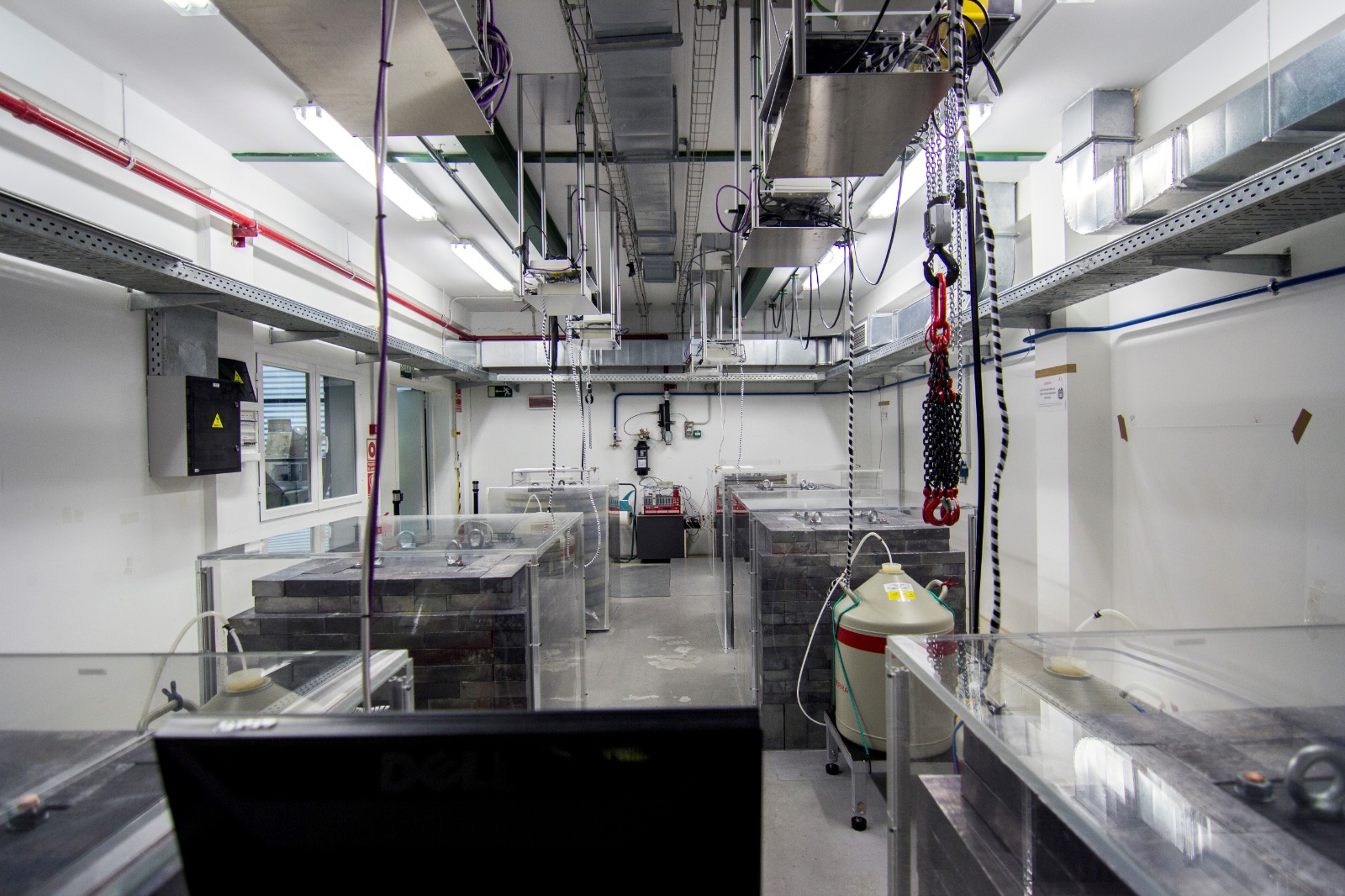}
\caption{\label{fig:ulbs}}
\end{subfigure}
\hfill
\begin{subfigure}{0.214\textwidth}
\centering
\includegraphics[width=\textwidth]{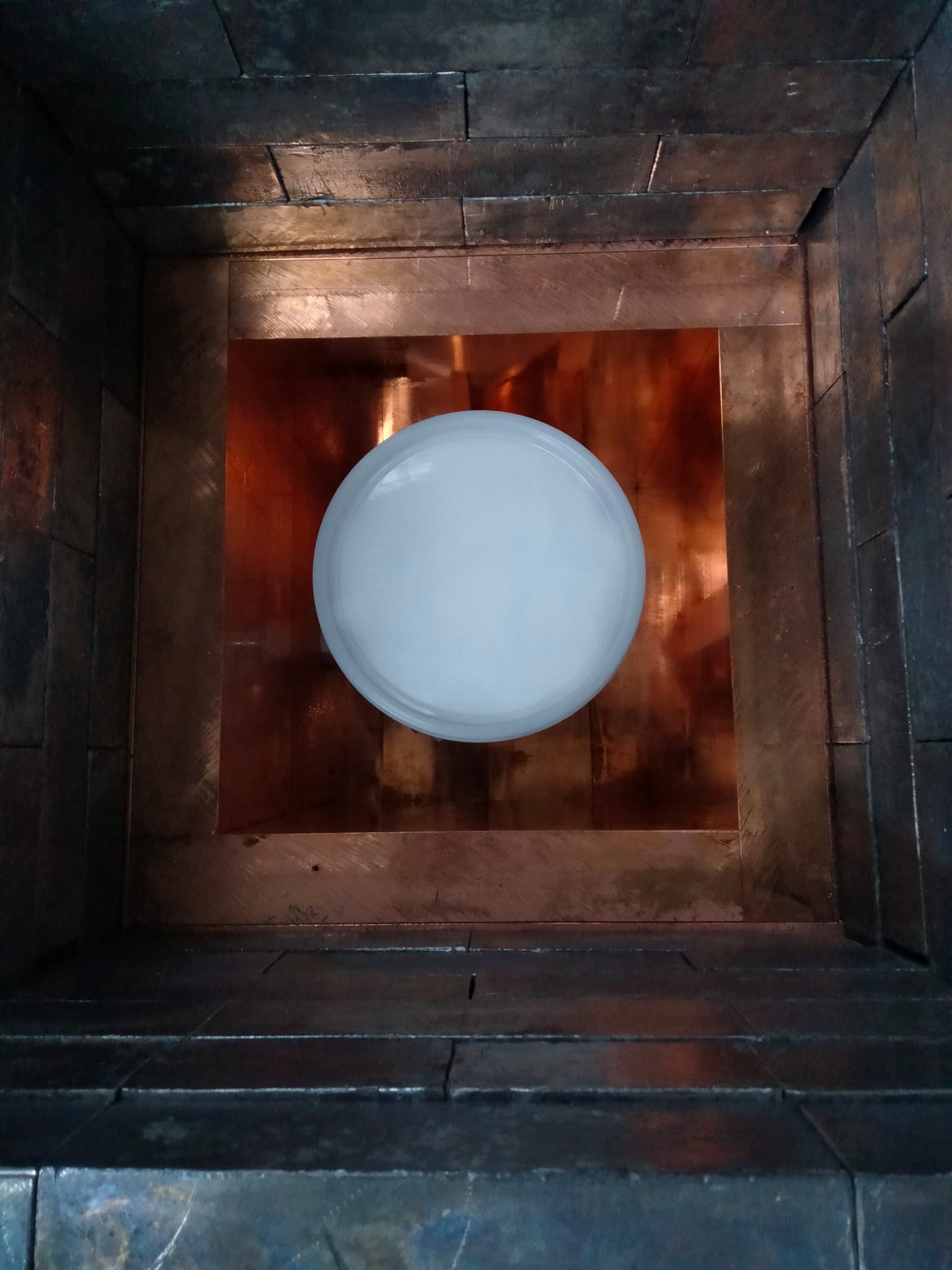}
\caption{\label{fig:inGeOroel}}
\end{subfigure}
\hfill
\begin{subfigure}{0.214\textwidth}
\centering
\includegraphics[width=\textwidth]{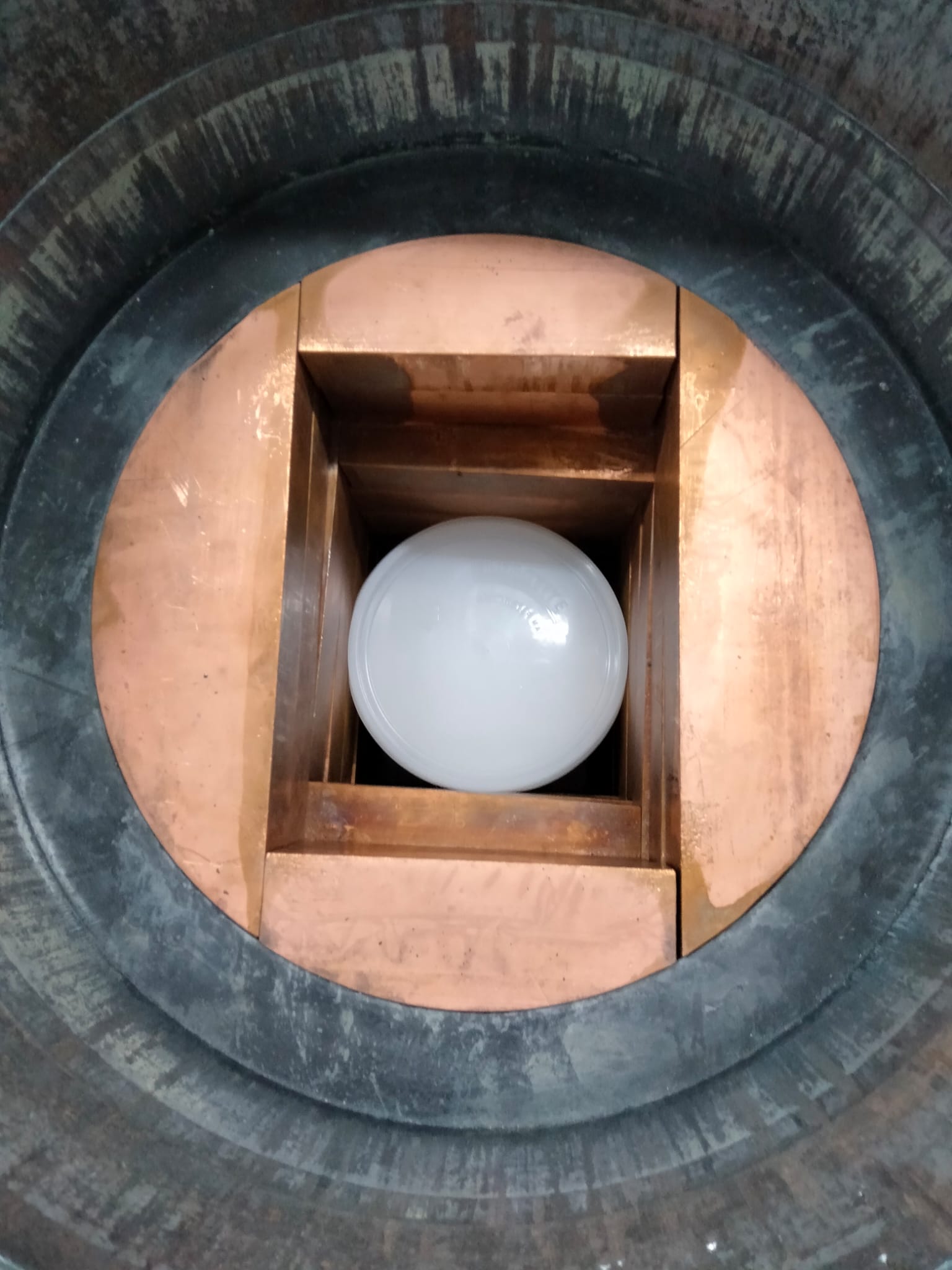}
\caption{\label{fig:inAsterix}}
\end{subfigure}
\caption{(a) ULBS laboratory in Hall C of LSC, (b) \GdSO~ sample inside GeOroel shield and (c) \GdSOw~sample inside Asterix shield.}
\end{figure}

The Ultra-Low Background Service (ULBS), working since 2010 in Hall C of LSC (Figure \ref{fig:ulbs}), offers a high-quality material assay facility to experiments. At present, it is equipped with six 2~kg p-type coaxial HPGe detectors with $\sim 100\%$ relative efficiency and two SAGe-Well High-Purity detectors, shielded with 20 cm of lead with a low contamination in $^{210}$Pb. An internal 10~cm layer of OFHC copper completes the shielding. For SK-Gd, two p-type detectors were mainly used: GeOroel and Asterix (see Table \ref{tab:Gedetectors}). 

Samples of $\sim$4~kg \GdSOw\ were packed in Marinelli beakers (Ga-Ma and Associates, Inc. model 445N-E). Measurements of around 30 days are necessary to reach mBq/kg sensitivity, with the condition that the radon contained in the air inside the HPGe detector shielding has decayed. To achieve this, each detector has an extra layer of shielding made of methacrylate which surrounds the lead shielding. To impede the exterior air (and the Rn) to enter inside the shielding, a slight over-pressure is created inside the copper shielding. The Cu shield inner volume available for sample measurement is $\sim 43$~L (see  Figures \ref{fig:inGeOroel} and \ref{fig:inAsterix}). This volume is flushed with $\sim 274$~L/h of a mixture of nitrogen and Rn-free air, which corresponds to about six full volume changes per hour. 
\subsubsection{Kamioka} %% Ichimura
Kamioka Observatory is located 2700~m water equivalent underground. In addition to housing the SK-Gd detector, several HPGe detectors are operated to support the many scientific goals of the laboratory. For the current \GdSOw\ assay program, the ultra-low background HPGe detector located in Lab-C, manufactured by Mirion Technologies France~\cite{CANBERRAFrance}, was used. Details of the shield geometry and detector performance can be found in~\cite{Abe_2020}.

Samples of \GdSOw\ are prepared in two ways. First, a molecular recognition resin embedded in the "Empore Radium Rad Disk"~\cite{10.1093/ptep/pty096,10.1093/ptep/ptaa105} was used to adsorb radium from the \GdSOw\ and increase its concentration. Since the size of the disk is small (47~mm diameter, 0.5~$\mu$m thickness), it could be laid directly onto the end cap of the HPGe detector to measure the concentrated radium activity in \GdSOw\ with high detection efficiency. A $^{226}$Ra activity of 0.5~mBq/kg is measurable with ten days of measurement using this method.

\begin{figure}[htp]
\begin{subfigure}{0.47\textwidth}
\centering
\includegraphics[width=\textwidth]{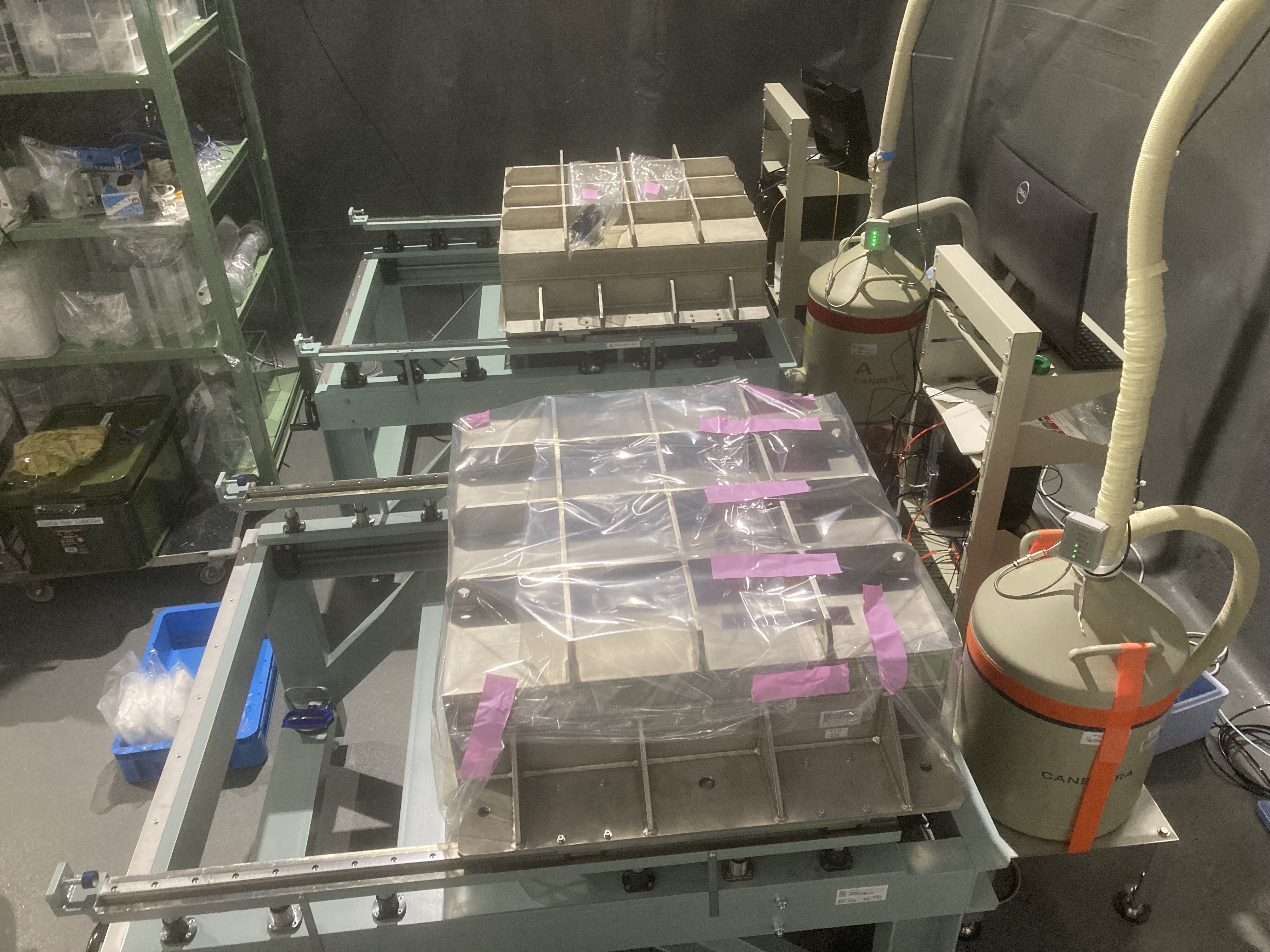}
\caption{\label{fig:LabC}}
\end{subfigure}
\hfill
\begin{subfigure}{0.47\textwidth}
\centering
\includegraphics[width=\textwidth]{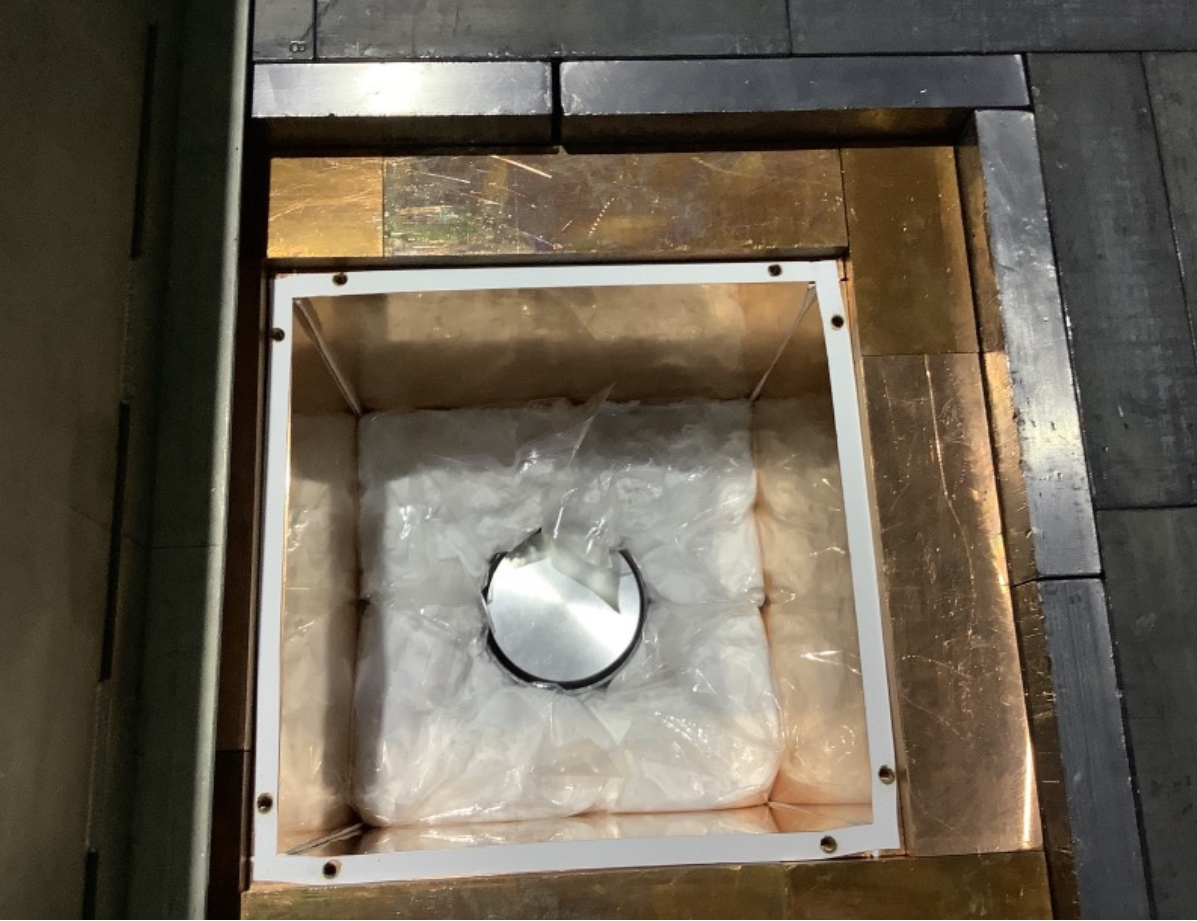}
\caption{\label{fig:LabCGePicture}}
\end{subfigure}
\caption{(a) Ge detectors in Lab-C at Kamioka Observatory, (b) Kamioka Lab-C Ge detector with 10~kg of \GdSOw\ in EVOH bags loaded for measurement.}
\end{figure}

Second, 2.5~kg of \GdSOw\ was packed into ethylene-vinyl alcohol (EVOH) bags. Any radon present was purged from the bags with radon-free air, the bags evacuated and then closed. Four such bags are loaded directly into the HPGe shield, surrounding the detector head and filling the space within the shield.
%This detector was used for the measurement of the $^{226}$Ra in the  gadolinium sulfate with chemical separation method\cite{10.1093/ptep/pty096}\cite{10.1093/ptep/ptaa105}.
%In this study, a large amount of gadolinium sulfate was measured without chemical separation in order to measure not only radium but also other potential RIs with the sensitivity required by SK-Gd.
%The sample chamber dimensions in the Lab-C HPGe shield is 23$\times$23$\times$48~cm. Subtracting the volume of the detector, this is enough available space to measure 10~kg of gadolinium sulfate at a time. 
Figure~\ref{fig:LabCGePicture} shows the detector with four EVOH bags of \GdSOw\ loaded for measurement. About two weeks of sample measurement time is required to achieve a MDA of $<0.5$~mBq/kg of $^{226}$Ra at 95\% confidence level.

%Figure~\ref{fig:GeSpectra} shows the energy spectrum of the sample measurement together with the background spectrum.
%Radio-activities were calculated from net count of sample data and background data at each characteristic-$\gamma$ energy (see example spectra in Fig.~\ref{fig:GeSpectra}),  after applying   branching ratio and detection efficiency. 

%For example, to evaluate the radioactivity of $^{214}$Bi, which is a daughter of $^{226}$Ra,  609 keV, 1120 keV and 1764 keV $\gamma$-ray were used.
A Monte Carlo simulation based on Geant4 was used to evaluate the detection efficiency for the sample geometries.

%About two weeks sample measurement can achieve the sensitivity of 0.5 mBq/kg of $^{226}$Ra.
%In addition to the Lab-C Ge detector, we use two HPGe detectors named "IPMU-P" and "IPMU-N" for screening at a level of a few mBq/kg of $^{226}$Ra using about 2.5 kg of gadolinium sulfate.  Details of these detectors are provided in~\cite{ABE2019171}. 

% == Fig. 

\subsection{HPGe Performance}

Table~\ref{tab:Gedetectors} shows relevant characteristics and some background count rates at relevant gamma energies of all the HPGe detectors used, as well as the number of SK-Gd samples assayed on each.

\begin{table}[htp]
%\centering
%\tiny
\small
\begin{tabular}{llllcccccc}
Lab &
  Detector &
  \begin{tabular}[c]{@{}l@{}}Mass \\ {[}kg{]}\end{tabular} &
  \begin{tabular}[c]{@{}l@{}}FWHM@\\ 1332 keV\\ {[}keV{]}\end{tabular} &
  \begin{tabular}[c]{@{}c@{}}COUNTS\\Integral\\ 60-2700\\ keV\end{tabular} &
  \begin{tabular}[c]{@{}c@{}} {[}/kg/day{]}\\$^{208}$Tl,\\ 2614\\ keV\end{tabular} &
  \begin{tabular}[c]{@{}c@{}}\\$^{214}$Bi,\\ 609\\ keV\end{tabular} &
  \begin{tabular}[c]{@{}c@{}}\\$^{60}$Co,\\ 1332\\ keV\end{tabular} &
  \begin{tabular}[c]{@{}c@{}}\\$^{40}$K,\\ 1461\\ keV\end{tabular} &
  \begin{tabular}[c]{@{}c@{}}SK-Gd \\ total \\ samples\end{tabular} \\
  \hline
BUGS   & Belmont   & 3.2  & 1.92 & 90.0  & 0.12 & 0.67 & 0.47 & 0.58 & 8  \\
BUGS   & Merrybent & 2.0  & 1.87 & 145.0 & 0.23 & 2.15 & 0.47 & 1.16 & 5  \\
LSC & GeOroel   & 2.31 & 2.22 & 128.7 & 0.53 & 0.89 & 0.06 & 0.46 & 3  \\
LSC & Asterix & 2.13 & 1.92 & 171.3& 0.11 & 1.10 & 0.28 & 0.61 & 13  \\
LSC & GeAnayet & 2.26 & 1.99 & 461.2 & 3.68 & 0.71 & 0.16 & 0.74 & 2 \\
Kamioka & Lab-C Ge & 1.68 & 2.39 & 104.5 & 0.08 & 0.39 & 0.41 & 0.44 & 23
\end{tabular}
\caption{HPGe detectors used and their main properties and background characteristics. The number of samples assayed by each detector is also shown, including batches measured but not selected for use in SK-Gd.}
\label{tab:Gedetectors}
\end{table}

\begin{figure}[htp]
    \centering 
    \includegraphics[width=1.0\textwidth]
    %{img/KamGeSpectra.eps}
    %{img/KamGeSpectra.v2.eps}
    %{img/KamGeSpectra.v3.eps}
    {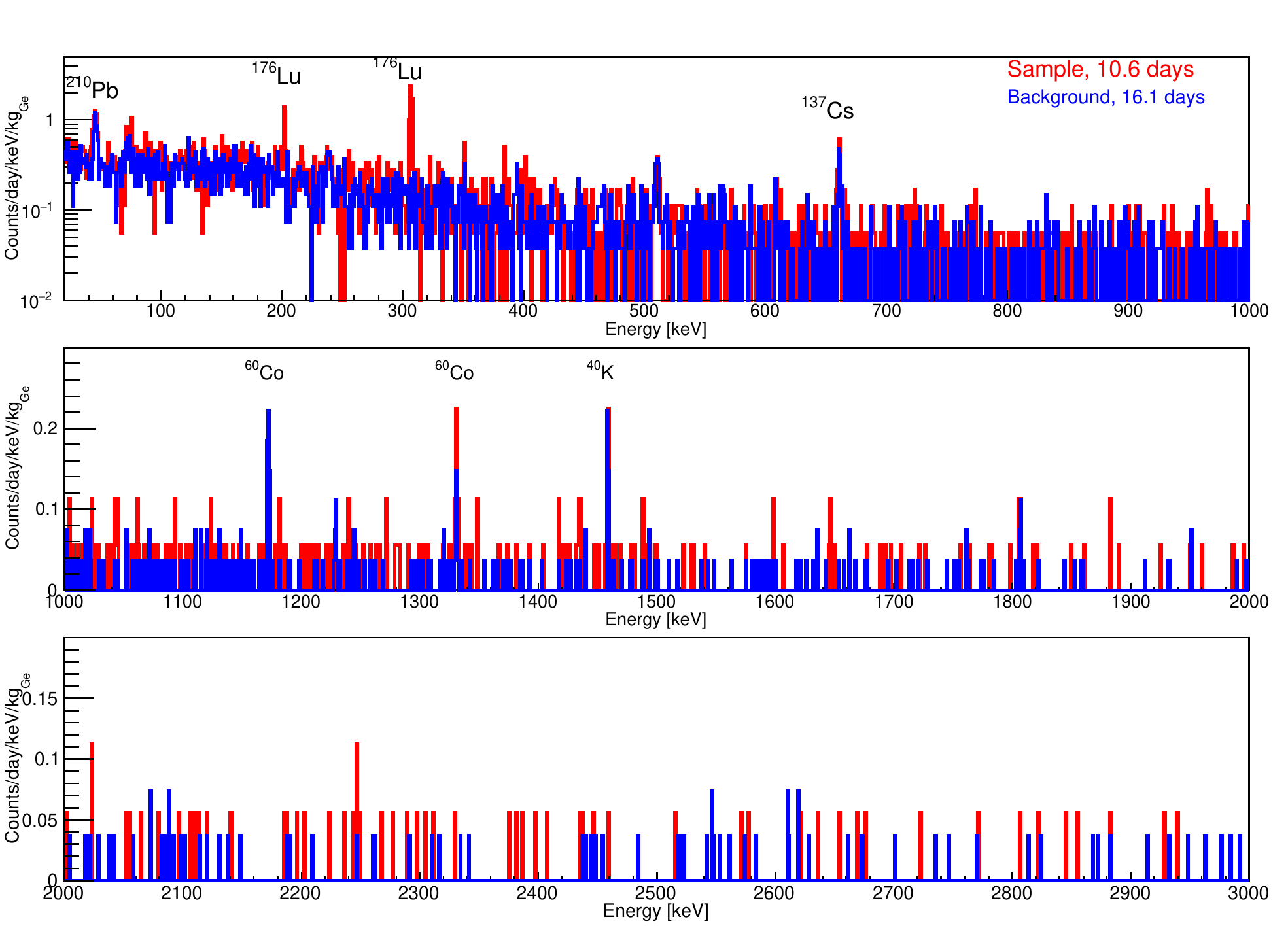}
    \caption{\label{fig:GeSpectra}
    A typical gamma spectrum (red) corresponding to 10~kg of \GdSOw\ measured by Kamioka's Lab-C Ge detector, and the detector background (blue). The 662~keV peak related to $^{137}$Cs arises from contamination in the shield materials. The $^{60}$Co peaks originate from cosmogenic neutron activation of the copper shield material.
    }
\end{figure}

Figure~\ref{fig:GeSpectra} shows a typical measurement spectrum (red) compared with the detector background (blue). The measurement corresponds to 10~kg of \GdSOw\ measured by Kamioka's Lab-C Ge detector. No peaks from daughters of $^{226}$Ra (for example, $^{214}$Pb: 295 keV, 352 keV, and $^{214}$Bi: 609 keV, 1120 keV, 1764 keV) are observed either in the sample or background data. The two peaks observed at 202 keV and 307 keV in the sample spectrum come from the decay of $^{176}$Lu, which is observed in most \GdSOw\ samples.

\begin{figure}
\begin{subfigure}{0.48\textwidth}
\centering
\includegraphics[width=\textwidth]{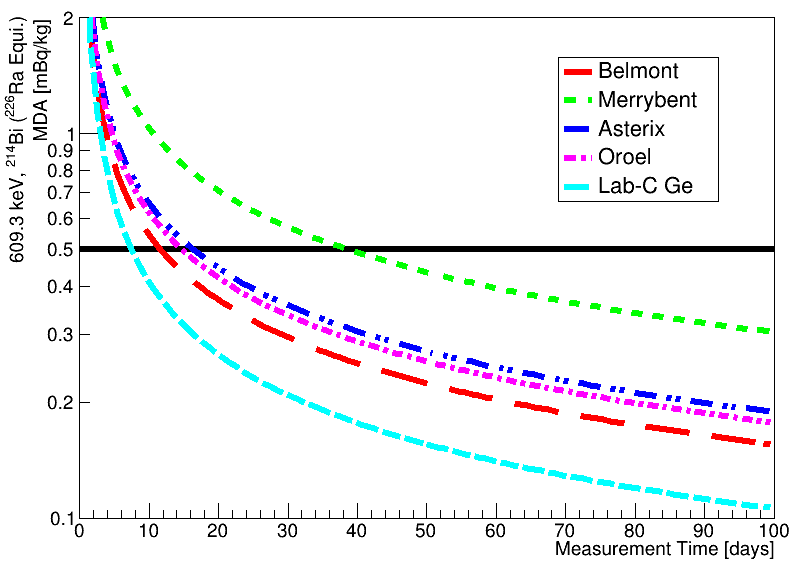}
\caption{\label{fig:sensitivity609}}
\end{subfigure}
\hfill
\begin{subfigure}{0.48\textwidth}
\centering
\includegraphics[width=\textwidth]{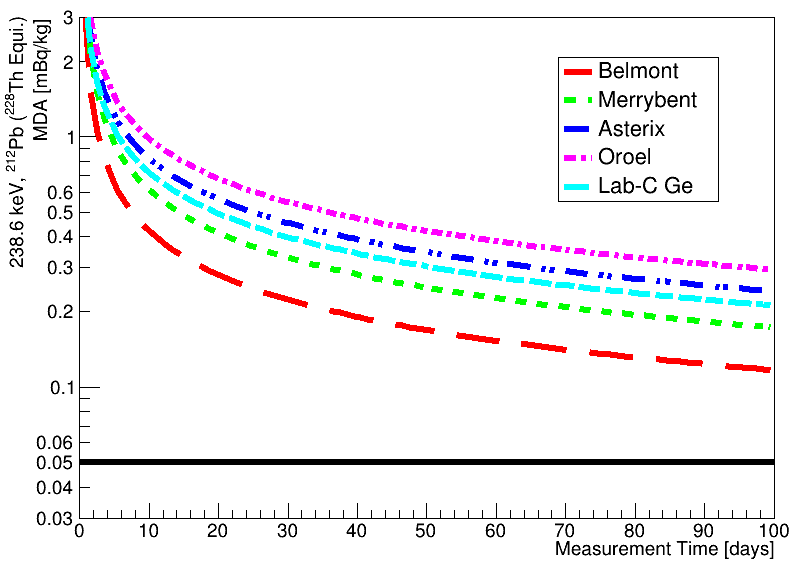}
\caption{\label{fig:sensitivity238}}
\end{subfigure}
\caption{\label{fig:MDA} The minimum detectable activity (MDA) for (a) the 609.3~keV gammas from $^{214}$Bi and (b) the 238.6~keV gammas from $^{212}$Pb for five HPGe detectors in this study.}
\end{figure}

A true comparison of the sensitivities of the HPGe detectors accounts for the detector backgrounds as well as the gamma detection efficiency for the particular sample. For \GdSOw\ samples prepared according to the particular procedures used in each laboratory, all of the detectors used in this study are sensitive to $<30$~mBq/kg of $^{235}$U and its daughters within just a few days of measurement. In contrast, none of the HPGe detectors are sensitive to $<0.05$~mBq/kg of $^{232}$Th daughters in any realistic measurement timescale (see Figure~\ref{fig:sensitivity238}). Since $^{238}$U is measurable using ICP-MS, achieving a sensitivity of $<5$~mBq/kg using HPGe methods is not critical. For $^{226}$Ra daughters, all HPGe detectors are sensitive to $<0.5$~mBq/kg within a few weeks of measurement (see Figure~\ref{fig:sensitivity609}). It is this requirement which primarily motivates the measurement time of \GdSOw\ samples on the HPGe detectors. 

\subsection{ICP-MS}\label{sec:icpmsintro}

In addition to HPGe gamma spectrometry at Kamioka Laboratory, specialty ICP-MS is used to assay U, Th, and Ce impurities.

To measure U and Th at the parts-per-trillion (ppt) level, a solid-phase extraction technique has been developed~\cite{10.1093/ptep/pty096}. First, a nitric acid aqueous solution in which a sample of \GdSOw\ is dissolved is passed through a well-washed chromatographic extraction resin, so that about 90\% or more U and Th are adsorbed on the resin but Gd is not. U and Th can then be eluted from the resin when a dilute nitric acid solution is passed through. Finally, by analyzing the elute with ICP-MS it is possible to measure trace amounts of U and Th without interference from Gd, which is reduced by a factor of about 10$^4$.

To assay Ce impurities, \GdSOw\ is diluted 10,000-fold by mass ratio with 2\% HNO$_3$.
Then the concentration of Ce in the aqueous solution is directly measured using ICP-MS. At this concentration, the matrix effect due to the existence of Gd is negligible since the Gd concentration is 0.01\%.

\section{Results}\label{sec:results} % Ikeda with help of Matt/Ichimura/H.Ito
Table~\ref{tab:ICPresult} shows the results from the ICP-MS assays of all batch samples of \GdSOw. All samples meet the criteria for U, Th and Ce contamination except for one which slightly exceeds the criteria value for Th. This will be discussed later in more detail. The mean Ce concentration in all approved samples of \GdSOw, weighted by the dissolved batch mass, is measured to be $11.1 \pm 3.8$~ppb, well below the 50~ppb requirement.

% == Fig. 
%\begin{figure}[htbp]
%    \centering 
%    \includegraphics[width=\textwidth]
%    {img/icpms_result.pdf}
%    {img/icpms_result_U_Th_Ce.pdf}

%    \caption{\label{fig:ICPresult}
%    Results from ICP-MS assays for all samples of \GdSOw\ approved for use in SK. The horizontal axis corresponds to the position of the sample in Table~\ref{tab:ICPresult}.}
%\end{figure}

\begin{table}[!ht]
%\tiny 
    \centering
    \begin{tabular}{r|ccc}
    \hline
        Batch ID & Ce [ppb] & U [ppt] & Th [ppt]  \\ 
         & $<$ 50 & $<$ 400 & $<$ 13  \\ \hline
%        17090X &  35±1.6 &  1.5±0.7 & 4.9±1.8  \\ \hline
%        180702 & $<$ 10 &  1.5±0.9 & 4.8±0.4  \\ \hline
%        180703 & $<$ 10  &  0.7±0.4 & 9.1±2.9  \\ \hline
%        190302 & 23.5±1.0 &  -3.7±3.9 &  -1.8±6.0  \\ \hline
%        190303 & 10.4±1.2 &  3.5±2.3 &  3.9±9.1  \\ \hline
%        190304 & 10.8±0.7 &  -4.7±7.9 &  7.1±7.5  \\ \hline
%        190502 & 10.0±0.6 &  1.6±2.0 &  -4.6±4.5  \\ \hline
%        190604 & 32.5±0.7 &  4.8±0.5 &  7.7±0.7  \\ \hline
%        190606 & 29.0±0.8 &  5.1±2.3 &  4.0±8.2  \\ \hline
%        190607 & 10.1±0.1 &  -1.5±5.9 &  11.6±2.3  \\ \hline
%        190608 & 13.8±1.0 &  -7.8±3.4 &  3.9±2.3  \\ \hline
%        190702 & 12.1±0.7 &  3.9±0.3 &  11.1±0.9  \\ \hline
%        190703 & 9.7±0.6 &  3.4±0.6 &  2.1±0.4  \\ \hline
%        190704 & 13.9±0.5 &  5.9±0.4 &  6.3±0.8  \\ \hline
%        190706 & 5.3±0.6 &  4.7±0.4 &  1.2±0.1  \\ \hline
%        190801 & 4.5±0.5 &  3.7±0.4 &  2.1±0.8  \\ \hline
%        190803 & 6.3±0.6 &  3.4±0.6 &  1.9±0.6  \\ \hline
%        190804 & 6.0±0.3 &  2.5±0.7 &  15.5±1.0  \\ \hline
%        190805 & 6.5±0.6 &  2.2±0.8 &  0.2±0.3  \\ \hline
%        190806 & 5.9±0.4 &  8.5±1.0 &  7.8±0.5  \\ \hline
%        190901 & 8.0±0.9 &  4.0±0.3 &  7.2±1.0  \\ \hline
%        190902 & 5.7±0.4 &  2.4±0.5 &  5.0±1.1  \\ \hline
%        190903 & 5.5±0.4 &  4.0±1.0 &  3.8±0.5  \\ \hline
%        190905 & 6.5±0.4 &  2.7±0.5 &  4.4±0.7  \\ \hline
%        200101 & 6.7±0.8 &  1.4±1.7 &  3.5±1.8  \\ \hline
%        200103 & 4.5±0.2 &  1.4±0.3 &  2.1±0.5  \\ \hline
%        200104 & 4.0±0.4 &  1.4±0.5 &  2.5±0.9  \\ \hline

        17090X &  35±1.6 &  $<$ 2.7 & 4.9±1.8  \\ \hline
        180702 & $<$ 10 &  $<$ 3.0 & 4.8±0.4  \\ \hline
        180703 & $<$ 10  &  $<$ 1.4 & 9.1±2.9  \\ \hline
        190302 & 23.5±1.0 &  $<$ 6.4 &  $<$ 9.8  \\ \hline
        190303 & 10.4±1.2 &  $<$ 7.3 &  $<$ 19  \\ \hline
        190304 & 10.8±0.7 &  $<$ 13 &  $<$ 19  \\ \hline
        190502 & 10.0±0.6 &  $<$ 4.9 &  $<$ 7.4  \\ \hline
        190604 & 32.5±0.7 &  4.8±0.5 &  7.7±0.7  \\ \hline
        190606 & 29.0±0.8 &  $<$ 8.9 &  $<$ 17  \\ \hline
        190607 & 10.1±0.1 &  $<$ 9.7 &  11.6±2.3  \\ \hline
        190608 & 13.8±1.0 &  $<$ 5.6 &  $<$ 7.7  \\ \hline
        190702 & 12.1±0.7 &  3.9±0.3 &  $<$ 12  \\ \hline
        190703 & 9.7±0.6 &  3.4±0.6 &  2.1±0.4  \\ \hline
        190704 & 13.9±0.5 &  5.9±0.4 &  6.3±0.8  \\ \hline
        190706 & 5.3±0.6 &  4.7±0.4 &  1.2±0.1  \\ \hline
        190801 & 4.5±0.5 &  3.7±0.4 &  2.1±0.8  \\ \hline
        190803 & 6.3±0.6 &  3.4±0.6 &  1.9±0.6  \\ \hline
        190804 & 6.0±0.3 &  2.5±0.7 &  15.5±1.0  \\ \hline
        190805 & 6.5±0.6 &  2.2±0.8 &  0.2±0.3  \\ \hline
        190806 & 5.9±0.4 &  8.5±1.0 &  7.8±0.5  \\ \hline
        190901 & 8.0±0.9 &  4.0±0.3 &  7.2±1.0  \\ \hline
        190902 & 5.7±0.4 &  2.4±0.5 &  5.0±1.1  \\ \hline
        190903 & 5.5±0.4 &  4.0±1.0 &  3.8±0.5  \\ \hline
        190905 & 6.5±0.4 &  2.7±0.5 &  4.4±0.7  \\ \hline
        200101 & 6.7±0.8 &  $<$ 4.2 &  $<$ 6.5  \\ \hline
        200103 & 4.5±0.2 &  1.4±0.3 &  2.1±0.5  \\ \hline
        200104 & 4.0±0.4 &  1.4±0.5 &  2.5±0.9  \\ \hline

    \end{tabular}
    \caption{\label{tab:ICPresult}
    Results from ICP-MS assays of all samples taken from batches which have been approved for use in SK-Gd. The SK requirement on the concentration of each element is indicated at the top of each column.}
\end{table}

Gamma spectrometric assay results for all SK-Gd batch samples are shown in Table~\ref{Tab:GeSummary}. For the late-chain $^{238}$U ($^{226}$Ra equilibrium) and entire $^{235}$U series, most of the limits set on the activity are below the detection limit for the given detector. Some activities in these sub-chains are reported with finite, non-zero values at the 95\% confidence level. The decision criteria for reporting one-sided or two-sided confidence intervals for low numbers of counts follows the procedure recommended in \cite{hurtgen2000}.

For $^{228}$Ra and $^{228}$Th, many of the later batches (after 190706) are observed to contain more than the criteria value of gamma activity. Considering the ICP-MS measurements of $^{232}$Th indicate an activity which satisfies the SK-Gd requirements, it is inferred that the production and processing of the \GdSOw\ powder disrupts the secular equilibrium in the decay chain, removing $^{232}$Th more efficiently than $^{228}$Ra. The significant excess of $^{228}$Ra daughters in these later batches is associated with a known excess of impurities in the raw material \GdOx\ which was used to produce these batches. As described in Section~\ref{sec:purification}, most lots of \GdOx\ were selected to contain 150-200~ppb of Th. The later lots contained more Th but could not be affordably replaced with a cleaner raw material. We justify the use of this material, despite the impurities, knowing that Ra is adsorbed by cation exchange resins in the SK-Gd water system~\cite{ABE2022166248}, and that the mean lifetime of $^{228}$Th is 2.7 years, well within the expected lifetime of the SK-Gd project.

We also note that many samples contained $^{176}$Lu and $^{138}$La. Since these nuclides have $Q$-values below 2~MeV, they are below the detection threshold in SK-Gd so do not meaningfully affect any physics analyses.

The total impurity activity budget for SK-Gd (Section~\ref{requirements}) is generally based on the requirement that backgrounds will not increase by more than 100\% of the pre-Gd backgrounds. For all of the limited decay chain activities, the radioactivity budget is shown in Table~\ref{tab:resultssummary} along with a summary of the estimated total impurities dissolved into SK-Gd via the \GdSOw. 

To estimate the total impurities in all batches, we first assume that each small (4-10~kg for HPGe measurements, several grams for ICP-MS measurements) assayed sample exactly represents the radioactivity of the full 500~kg batch from which it came, then combine the finite activities and upper limits in two ways. First, a lower bound on the total activity is the sum of all finite measured (FM) activities, with the errors combined in quadrature. Second, a conservative upper bound on the total activity is the sum of all 95\% confidence level upper limits (UL), which includes the reported one-sided confidence intervals as well as the finite measurements plus 1.645 times the measurement standard deviation.

For $^{238}$U and $^{232}$Th, the ICP-MS measurements show that the total added radioactivity is well below the SK-Gd budget. The HPGe results for all parts of the $^{238}$U and $^{235}$U decay chains show that, even in the worst case scenario, the SK-Gd requirements are met even though some samples of \GdSOw\ contain activities which exceed the required specific activity for these nuclides. 

For $^{228}$Ra and $^{228}$Th, the sum of finite measurements (FM) in all batch samples is more than half of the total SK-Gd budget, even though these dissolved batches only consist of about 10\% of the design Gd concentration. The conservative upper limit (UL) for these decay chain activities is 30\% to 60\% greater than the total SK-Gd budget. Without better measurements of the $^{228}$Ra and $^{228}$Th activities, it is not known whether the SK-Gd total budget is met or exceeded.

Since the $^{232}$Th activity is shown to be well within the total SK-Gd budget, and since it is known that Ra is adsorbed by the resin in the SK-Gd water system, only the observed activity of $^{228}$Th is expected to be problematic for SK-Gd. The HPGe result for $^{228}$Th represents the true activity of $^{208}$Tl at the time of measurement, which will decay with a 1.9~yr half-life assuming all $^{228}$Ra is removed during the SK-Gd water filtration. By one half-life, the worst-case estimate of total $^{208}$Tl activity is within the SK-Gd requirement.

% \usepackage{multirow}
\begin{landscape}
\begin{table}
\tiny 
\renewcommand{\arraystretch}{1.2}
\begin{tabular}{ccc|ccccccccccc}
\hline
\multicolumn{1}{|c|}{\multirow[t]{4}{*}{}}     & \multicolumn{1}{c|}{\multirow[t]{4}{*}{}} & \multirow[t]{2}{*}{}                        & \multicolumn{11}{c|}{}  \\[-2.2ex]
\multicolumn{1}{|c|}{\multirow[t]{4}{*}{Sample}}     & \multicolumn{1}{c|}{\multirow[t]{4}{*}{Laboratory}} & \multirow[t]{2}{*}{Detector / Method}                        & \multicolumn{11}{c|}{Activity (mBq/kg, 95\% c.l.)}  \\ \cline{4-14} 
\multicolumn{1}{|c|}{}                        & \multicolumn{1}{c|}{}                      &                                                  & \multicolumn{2}{c|}{}                                                    & \multicolumn{2}{c|}{}                                                  & \multicolumn{2}{c|}{}                                                   & \multicolumn{1}{c|}{\multirow{3}{*}{}} & \multicolumn{1}{c|}{\multirow{3}{*}{}} & \multicolumn{1}{c|}{\multirow{3}{*}{}} & \multicolumn{1}{c|}{\multirow{3}{*}{}} & \multicolumn{1}{c|}{\multirow{3}{*}{}} \\[-2.2ex] 
\multicolumn{1}{|c|}{}                        & \multicolumn{1}{c|}{}                      &                                                  & \multicolumn{2}{c|}{$^{238}$U Chain}                                                    & \multicolumn{2}{c|}{$^{232}$Th Chain}                                                  & \multicolumn{2}{c|}{$^{235}$U Chain}                                                   & \multicolumn{1}{c|}{\multirow{3}{*}{$^{40}$K}} & \multicolumn{1}{c|}{\multirow{3}{*}{$^{138}$La}} & \multicolumn{1}{c|}{\multirow{3}{*}{$^{176}$Lu}} & \multicolumn{1}{c|}{\multirow{3}{*}{$^{134}$Cs}} & \multicolumn{1}{c|}{\multirow{3}{*}{$^{137}$Cs}} \\ \cline{4-9}
\multicolumn{1}{|c|}{}                        & \multicolumn{1}{c|}{}                      &                                                  & \multicolumn{1}{c|}{}            & \multicolumn{1}{c|}{}           & \multicolumn{1}{c|}{}           & \multicolumn{1}{c|}{}           & \multicolumn{1}{c|}{}           & \multicolumn{1}{c|}{}           & \multicolumn{1}{c|}{}                     & \multicolumn{1}{c|}{}                       & \multicolumn{1}{c|}{}                       & \multicolumn{1}{c|}{}                       & \multicolumn{1}{c|}{}                       \\[-2.2ex]
\multicolumn{1}{|c|}{}                        & \multicolumn{1}{c|}{}                      &                                                  & \multicolumn{1}{c|}{E, $^{238}$U eq.}            & \multicolumn{1}{c|}{L, $^{226}$Ra eq.}           & \multicolumn{1}{c|}{E, $^{228}$Ra eq.}           & \multicolumn{1}{c|}{L, $^{228}$Th eq.}           & \multicolumn{1}{c|}{E, $^{235}$U eq.}           & \multicolumn{1}{c|}{L, $^{227}$Ac eq.}           & \multicolumn{1}{c|}{}                     & \multicolumn{1}{c|}{}                       & \multicolumn{1}{c|}{}                       & \multicolumn{1}{c|}{}                       & \multicolumn{1}{c|}{}                       \\ \cline{4-14}
\multicolumn{1}{|c|}{}                        & \multicolumn{1}{c|}{}                      & \multicolumn{1}{r|}{} & \multicolumn{1}{c|}{}      & \multicolumn{1}{c|}{}   & \multicolumn{1}{c|}{}  & \multicolumn{1}{c|}{}            & \multicolumn{1}{c|}{}    & \multicolumn{1}{c|}{}    & \multicolumn{1}{c|}{}                     & \multicolumn{1}{c|}{}                       & \multicolumn{1}{c|}{}                       & \multicolumn{1}{c|}{}                       & \multicolumn{1}{c|}{}                       \\[-2.2ex]
\multicolumn{1}{|c|}{}                        & \multicolumn{1}{c|}{}                      & \multicolumn{1}{r|}{SK-Gd Req. $\rightarrow$} & \multicolumn{1}{c|}{\textless 5}      & \multicolumn{1}{c|}{\textless 0.5}   & \multicolumn{1}{c|}{\textless 0.05}  & \multicolumn{1}{c|}{\textless 0.05}            & \multicolumn{1}{c|}{\textless 30}    & \multicolumn{1}{c|}{\textless 30}    & \multicolumn{1}{c|}{-}                     & \multicolumn{1}{c|}{-}                       & \multicolumn{1}{c|}{-}                       & \multicolumn{1}{c|}{-}                       & \multicolumn{1}{c|}{-}                       \\ \hline\hline
\multicolumn{1}{|c|}{17090X}                  & \multicolumn{1}{c|}{Canfranc}              & ge-Asterix                                       & \multicolumn{1}{c|}{\textless 6}    & \multicolumn{1}{c|}{\textless 0.21}  & \multicolumn{1}{c|}{\textless 0.30}  & \multicolumn{1}{c|}{\textless 0.30}  & \multicolumn{1}{c|}{\textless 0.42}  & \multicolumn{1}{c|}{\textless 1.6}  & \multicolumn{1}{c|}{\textless 1.0}        & \multicolumn{1}{c|}{\textless 0.14}               & \multicolumn{1}{c|}{\textless 0.13±0.03}         & \multicolumn{1}{c|}{\textless 0.07}         & \multicolumn{1}{c|}{\textless 0.13}         \\ \hline
\multicolumn{1}{|c|}{180702}                  & \multicolumn{1}{c|}{Canfranc}              & ge-Asterix                                       & \multicolumn{1}{c|}{\textless 3.1}    & \multicolumn{1}{c|}{\textless 0.12}  & \multicolumn{1}{c|}{\textless 0.22}  & \multicolumn{1}{c|}{\textless 0.21}  & \multicolumn{1}{c|}{\textless 0.3}  & \multicolumn{1}{c|}{\textless 1.1}  & \multicolumn{1}{c|}{\textless 0.5}        & \multicolumn{1}{c|}{0.13±0.04}               & \multicolumn{1}{c|}{\textless 0.24±0.03}         & \multicolumn{1}{c|}{\textless 0.07}         & \multicolumn{1}{c|}{\textless 0.08}         \\ \hline
\multicolumn{1}{|c|}{180703}                  & \multicolumn{1}{c|}{Canfranc}              & ge-Asterix                                       & \multicolumn{1}{c|}{\textless 4.5}    & \multicolumn{1}{c|}{\textless 0.24}  & \multicolumn{1}{c|}{\textless 0.44}  & \multicolumn{1}{c|}{\textless 0.38}  & \multicolumn{1}{c|}{\textless 0.3}  & \multicolumn{1}{c|}{\textless 1.1}  & \multicolumn{1}{c|}{\textless 0.5}        & \multicolumn{1}{c|}{\textless 0.14}               & \multicolumn{1}{c|}{\textless 0.22±0.03}         & \multicolumn{1}{c|}{\textless 0.07}         & \multicolumn{1}{c|}{\textless 0.07}         \\ \hline
\multicolumn{1}{|c|}{190302}                  & \multicolumn{1}{c|}{Canfranc}              & ge-Asterix                                       & \multicolumn{1}{c|}{\textless 4.9}    & \multicolumn{1}{c|}{\textless 0.32}  & \multicolumn{1}{c|}{\textless 0.35}  & \multicolumn{1}{c|}{\textless 0.29}  & \multicolumn{1}{c|}{\textless 0.42}  & \multicolumn{1}{c|}{\textless 0.92}  & \multicolumn{1}{c|}{\textless 1.6}        & \multicolumn{1}{c|}{0.26±0.1}               & \multicolumn{1}{c|}{\textless 0.21}         & \multicolumn{1}{c|}{\textless 0.09}         & \multicolumn{1}{c|}{\textless 0.09}         \\ \hline
\multicolumn{1}{|c|}{190303}                  & \multicolumn{1}{c|}{Canfranc}              & ge-Asterix                                       & \multicolumn{1}{c|}{\textless 4.2}    & \multicolumn{1}{c|}{\textless 0.3}   & \multicolumn{1}{c|}{\textless 0.44}  & \multicolumn{1}{c|}{\textless 0.29}  & \multicolumn{1}{c|}{\textless 0.39}  & \multicolumn{1}{c|}{\textless 0.81}  & \multicolumn{1}{c|}{\textless 1.5}        & \multicolumn{1}{c|}{0.45±0.09}              & \multicolumn{1}{c|}{0.16±0.12}              & \multicolumn{1}{c|}{\textless 0.08}         & \multicolumn{1}{c|}{\textless 0.09}         \\ \hline
\multicolumn{1}{|c|}{190304}                  & \multicolumn{1}{c|}{Canfranc}              & ge-Asterix                                            & \multicolumn{1}{c|}{\textless 5.5}     & \multicolumn{1}{c|}{\textless 0.42}   & \multicolumn{1}{c|}{\textless 0.55}   & \multicolumn{1}{c|}{\textless 0.36}   & \multicolumn{1}{c|}{\textless 0.52}   & \multicolumn{1}{c|}{\textless 1.22}   & \multicolumn{1}{c|}{\textless 2.1}         & \multicolumn{1}{c|}{0.40±0.11}              & \multicolumn{1}{c|}{\textless 0.21}         & \multicolumn{1}{c|}{\textless 0.13}                      & \multicolumn{1}{c|}{\textless 0.14}         \\ \hline
\multicolumn{1}{|c|}{\multirow{2}{*}{190502}} & \multicolumn{1}{c|}{Boulby}                & Belmont                                       & \multicolumn{1}{c|}{\textless 5.4}    & \multicolumn{1}{c|}{\textless 0.49}   & \multicolumn{1}{c|}{\textless 0.95}  & \multicolumn{1}{c|}{\textless 0.48}  & \multicolumn{1}{c|}{\textless 0.36}  & \multicolumn{1}{c|}{\textless 1.7}   & \multicolumn{1}{c|}{\textless 2.8}        & \multicolumn{1}{c|}{\textless 0.28}         & \multicolumn{1}{c|}{0.49±0.08}              & \multicolumn{1}{c|}{-}                      & \multicolumn{1}{c|}{\textless 0.10}         \\ \cline{2-14} 
\multicolumn{1}{|c|}{}                        & \multicolumn{1}{c|}{Kamioka}               & Lab-C Ge                                                 & \multicolumn{1}{c|}{\textless 25.0}   & \multicolumn{1}{c|}{\textless 0.75}  & \multicolumn{1}{c|}{\textless 0.52}  & \multicolumn{1}{c|}{\textless 0.36}  & \multicolumn{1}{c|}{\textless 9}     & \multicolumn{1}{c|}{7.9±0.8}         & \multicolumn{1}{c|}{ \textless 1.63}                    & \multicolumn{1}{c|}{\textless 0.37}                      & \multicolumn{1}{c|}{0.68±0.18}              & \multicolumn{1}{c|}{\textless 0.16}                      & \multicolumn{1}{c|}{\textless 0.22}                      \\ \hline
\multicolumn{1}{|c|}{\multirow{2}{*}{190604}} & \multicolumn{1}{c|}{Boulby}                & Belmont                                           & \multicolumn{1}{c|}{\textless{}9.80}  & \multicolumn{1}{c|}{\textless{}0.47} & \multicolumn{1}{c|}{\textless{}0.61} & \multicolumn{1}{c|}{\textless{}0.50} & \multicolumn{1}{c|}{\textless{}0.45} & \multicolumn{1}{c|}{\textless{}2.33} & \multicolumn{1}{c|}{\textless{}2.45}      & \multicolumn{1}{c|}{\textless{}0.21}        & \multicolumn{1}{c|}{0.97±0.11}              & \multicolumn{1}{c|}{-}                      & \multicolumn{1}{c|}{\textless{}0.08}       \\ \cline{2-14} 
\multicolumn{1}{|c|}{}                        & \multicolumn{1}{c|}{Kamioka}               & Lab-C Ge                                                 & \multicolumn{1}{c|}{\textless 26.9}   & \multicolumn{1}{c|}{\textless 0.68}  & \multicolumn{1}{c|}{\textless 0.55}  & \multicolumn{1}{c|}{\textless 0.33}  & \multicolumn{1}{c|}{\textless 4.6}   & \multicolumn{1}{c|}{\textless 1.2}             & \multicolumn{1}{c|}{\textless 2.02}                    & \multicolumn{1}{c|}{\textless 0.36}                      & \multicolumn{1}{c|}{1.43±0.19}              & \multicolumn{1}{c|}{\textless 0.19}                      & \multicolumn{1}{c|}{\textless 0.34}                      \\ \hline
%\multicolumn{1}{|c|}{}                        & \multicolumn{1}{c|}{Kamioka}               & Lab-C Ge 90\% CL.                                                & \multicolumn{1}{c|}{\textless 23.1}   & \multicolumn{1}{c|}{\textless 0.60}  & \multicolumn{1}{c|}{\textless 0.43}  & \multicolumn{1}{c|}{\textless 0.26}  & \multicolumn{1}{c|}{\textless 3.6}   & \multicolumn{1}{c|}{1.2}             & \multicolumn{1}{c|}{-}                    & \multicolumn{1}{c|}{-}                      & \multicolumn{1}{c|}{1.43±0.19}              & \multicolumn{1}{c|}{-}                      & \multicolumn{1}{c|}{-}                      \\ \hline
\multicolumn{1}{|c|}{\multirow{3}{*}{190606}} & \multicolumn{1}{c|}{Boulby}                & Merrybent                      & \multicolumn{1}{c|}{\textless 13.1}   & \multicolumn{1}{c|}{\textless 0.84}  & \multicolumn{1}{c|}{\textless 0.79}  & \multicolumn{1}{c|}{\textless 0.63}  & \multicolumn{1}{c|}{\textless 0.37}  & \multicolumn{1}{c|}{2.6±0.6}         & \multicolumn{1}{c|}{\textless 3.27}       & \multicolumn{1}{c|}{\textless 0.29}         & \multicolumn{1}{c|}{1.23±0.16}              & \multicolumn{1}{c|}{-}                      & \multicolumn{1}{c|}{\textless 0.13}         \\ \cline{2-14} 
\multicolumn{1}{|c|}{}                        & \multicolumn{1}{c|}{Kamioka}               & Lab-C Ge                                                 & \multicolumn{1}{c|}{\textless 17.3}   & \multicolumn{1}{c|}{\textless 1.36}       & \multicolumn{1}{c|}{\textless 0.91}  & \multicolumn{1}{c|}{\textless 0.94}  & \multicolumn{1}{c|}{\textless 8.3}   & \multicolumn{1}{c|}{2.6±1.3}         & \multicolumn{1}{c|}{\textless 3.20}                    & \multicolumn{1}{c|}{\textless 0.26}                      & \multicolumn{1}{c|}{0.74±0.29}              & \multicolumn{1}{c|}{\textless 0.39}                      & \multicolumn{1}{c|}{\textless 0.50}                      \\ \cline{2-14} 
%\multicolumn{1}{|c|}{}                        & \multicolumn{1}{c|}{Kamioka}               & Lab-C Ge                                                 & \multicolumn{1}{c|}{\textless 13.5}   & \multicolumn{1}{c|}{1.04±0.38}       & \multicolumn{1}{c|}{\textless 0.71}  & \multicolumn{1}{c|}{\textless 0.82}  & \multicolumn{1}{c|}{\textless 6.5}   & \multicolumn{1}{c|}{2.7±1.2}         & \multicolumn{1}{c|}{-}                    & \multicolumn{1}{c|}{-}                      & \multicolumn{1}{c|}{0.74±0.29}              & \multicolumn{1}{c|}{-}                      & \multicolumn{1}{c|}{-}                      \\ \cline{2-14} 
\multicolumn{1}{|c|}{}                        & \multicolumn{1}{c|}{Kamioka}               & Lab-C Ge, Ra Disk                                          & \multicolumn{1}{c|}{-}                 & \multicolumn{1}{c|}{\textless 0.31}  & \multicolumn{1}{c|}{\textless 0.82}  & \multicolumn{1}{c|}{\textless 0.48}  & \multicolumn{1}{c|}{-}                & \multicolumn{1}{c|}{-}                & \multicolumn{1}{c|}{-}                     & \multicolumn{1}{c|}{-}                       & \multicolumn{1}{c|}{-}                       & \multicolumn{1}{c|}{-}                       & \multicolumn{1}{c|}{-}                       \\ \hline
%\multicolumn{1}{|c|}{}                        & \multicolumn{1}{c|}{Kamioka}               & Lab-C Ge, Ra Disk                                          & \multicolumn{1}{c|}{-}                 & \multicolumn{1}{c|}{\textless 0.24}  & \multicolumn{1}{c|}{\textless 0.71}  & \multicolumn{1}{c|}{\textless 0.40}  & \multicolumn{1}{c|}{-}                & \multicolumn{1}{c|}{-}                & \multicolumn{1}{c|}{-}                     & \multicolumn{1}{c|}{-}                       & \multicolumn{1}{c|}{-}                       & \multicolumn{1}{c|}{-}                       & \multicolumn{1}{c|}{-}                       \\ \hline
\multicolumn{1}{|c|}{190607}                  & \multicolumn{1}{c|}{Canfranc}              & ge-Oroel                                         & \multicolumn{1}{c|}{\textless 3.6}    & \multicolumn{1}{c|}{\textless 0.30}  & \multicolumn{1}{c|}{\textless 0.79}  & \multicolumn{1}{c|}{\textless 0.42}  & \multicolumn{1}{c|}{\textless 0.30}  & \multicolumn{1}{c|}{\textless 0.96}  & \multicolumn{1}{c|}{\textless 1.59}       & \multicolumn{1}{c|}{\textless 0.18}         & \multicolumn{1}{c|}{\textless 0.13}         & \multicolumn{1}{c|}{\textless 0.12}         & \multicolumn{1}{c|}{\textless{}0.09}        \\ \hline
\multicolumn{1}{|c|}{\multirow{3}{*}{190608}} & \multicolumn{1}{c|}{Canfranc}              & ge-Asterix                                       & \multicolumn{1}{c|}{\textless 4.4}    & \multicolumn{1}{c|}{\textless 0.53}  & \multicolumn{1}{c|}{\textless 0.43}  & \multicolumn{1}{c|}{\textless 0.35}  & \multicolumn{1}{c|}{\textless 0.40}  & \multicolumn{1}{c|}{\textless 0.88}  & \multicolumn{1}{c|}{\textless 1.50}       & \multicolumn{1}{c|}{\textless 0.14}         & \multicolumn{1}{c|}{\textless 0.25}         & \multicolumn{1}{c|}{\textless 0.08}         & \multicolumn{1}{c|}{\textless 0.09}         \\ \cline{2-14} 
\multicolumn{1}{|c|}{}                        & \multicolumn{1}{c|}{Kamioka}               & Lab-C Ge                                                 & \multicolumn{1}{c|}{\textless 23.2}   & \multicolumn{1}{c|}{\textless 1.06}       & \multicolumn{1}{c|}{\textless 1.38}  & \multicolumn{1}{c|}{\textless 0.80}  & \multicolumn{1}{c|}{\textless 4.3}   & \multicolumn{1}{c|}{\textless 1.8}             & \multicolumn{1}{c|}{\textless 2.15}                    & \multicolumn{1}{c|}{\textless 0.49}                      & \multicolumn{1}{c|}{\textless 0.51}         & \multicolumn{1}{c|}{\textless 0.21}                      & \multicolumn{1}{c|}{\textless 0.30}                      \\ \cline{2-14} 
%\multicolumn{1}{|c|}{}                        & \multicolumn{1}{c|}{Kamioka}               & Lab-C Ge                                                 & \multicolumn{1}{c|}{\textless 20.4}   & \multicolumn{1}{c|}{0.99±0.30}       & \multicolumn{1}{c|}{\textless 1.22}  & \multicolumn{1}{c|}{\textless 0.71}  & \multicolumn{1}{c|}{\textless 3.4}   & \multicolumn{1}{c|}{1.6}             & \multicolumn{1}{c|}{-}                    & \multicolumn{1}{c|}{-}                      & \multicolumn{1}{c|}{\textless 0.45}         & \multicolumn{1}{c|}{-}                      & \multicolumn{1}{c|}{-}                      \\ \cline{2-14} 
\multicolumn{1}{|c|}{}                        & \multicolumn{1}{c|}{Kamioka}               & Lab-C Ge, Ra Disk                                          & \multicolumn{1}{c|}{-}                 & \multicolumn{1}{c|}{\textless 0.63}  & \multicolumn{1}{c|}{\textless 0.52}  & \multicolumn{1}{c|}{\textless 0.61}  & \multicolumn{1}{c|}{-}                & \multicolumn{1}{c|}{-}                & \multicolumn{1}{c|}{-}                     & \multicolumn{1}{c|}{-}                       & \multicolumn{1}{c|}{-}                       & \multicolumn{1}{c|}{-}                       & \multicolumn{1}{c|}{-}                       \\ \hline
%\multicolumn{1}{|c|}{}                        & \multicolumn{1}{c|}{Kamioka}               & Lab-C Ge, Ra Disk                                          & \multicolumn{1}{c|}{-}                 & \multicolumn{1}{c|}{\textless 0.49}  & \multicolumn{1}{c|}{\textless 0.43}  & \multicolumn{1}{c|}{\textless 0.55}  & \multicolumn{1}{c|}{-}                & \multicolumn{1}{c|}{-}                & \multicolumn{1}{c|}{-}                     & \multicolumn{1}{c|}{-}                       & \multicolumn{1}{c|}{-}                       & \multicolumn{1}{c|}{-}                       & \multicolumn{1}{c|}{-}                       \\ \hline
\multicolumn{1}{|c|}{\multirow{2}{*}{190702}} & \multicolumn{1}{c|}{Canfranc}              & ge-Oroel                                         & \multicolumn{1}{c|}{\textless 5.5}   & \multicolumn{1}{c|}{\textless 0.45}  & \multicolumn{1}{c|}{\textless 1.11}  & \multicolumn{1}{c|}{\textless 0.50}  & \multicolumn{1}{c|}{\textless 0.37}  & \multicolumn{1}{c|}{2.4±0.9}         & \multicolumn{1}{c|}{\textless 1.5}        & \multicolumn{1}{c|}{\textless 0.20}         & \multicolumn{1}{c|}{0.23±0.13}              & \multicolumn{1}{c|}{\textless 0.12}         & \multicolumn{1}{c|}{\textless 0.11}         \\ \cline{2-14} 
\multicolumn{1}{|c|}{}                        & \multicolumn{1}{c|}{Kamioka}               & Lab-C Ge                                     & \multicolumn{1}{c|}{\textless 12.0}   & \multicolumn{1}{c|}{\textless 0.63}  & \multicolumn{1}{c|}{\textless 1.08}  & \multicolumn{1}{c|}{\textless 0.33}  & \multicolumn{1}{c|}{\textless 3.4}   & \multicolumn{1}{c|}{\textless 1.6}   & \multicolumn{1}{c|}{ \textless 1.99}                   & \multicolumn{1}{c|}{\textless 0.28}                      & \multicolumn{1}{c|}{0.28±0.12}         & \multicolumn{1}{c|}{\textless 0.17}                      & \multicolumn{1}{c|}{\textless 0.28}                      \\ \hline
%\multicolumn{1}{|c|}{}                        & \multicolumn{1}{c|}{Kamioka}               & Lab-C Ge                                     & \multicolumn{1}{c|}{\textless 11.4}   & \multicolumn{1}{c|}{\textless 0.55}  & \multicolumn{1}{c|}{\textless 1.09}  & \multicolumn{1}{c|}{\textless 0.30}  & \multicolumn{1}{c|}{\textless 3.0}   & \multicolumn{1}{c|}{\textless 1.5}   & \multicolumn{1}{c|}{-}                    & \multicolumn{1}{c|}{-}                      & \multicolumn{1}{c|}{\textless 0.35}         & \multicolumn{1}{c|}{-}                      & \multicolumn{1}{c|}{-}                      \\ \hline
\multicolumn{1}{|c|}{190703}                  & \multicolumn{1}{c|}{Canfranc}              & ge-Asterix                                       & \multicolumn{1}{c|}{\textless 4.2}    & \multicolumn{1}{c|}{\textless 0.35}  & \multicolumn{1}{c|}{\textless 0.51}  & \multicolumn{1}{c|}{\textless 0.50}  & \multicolumn{1}{c|}{\textless 0.45}  & \multicolumn{1}{c|}{1.8±1.0}         & \multicolumn{1}{c|}{\textless 1.7}        & \multicolumn{1}{c|}{\textless 0.20}         & \multicolumn{1}{c|}{0.51±0.13}              & \multicolumn{1}{c|}{\textless 0.10}         & \multicolumn{1}{c|}{\textless 0.10}         \\ \hline
\multicolumn{1}{|c|}{190704}                  & \multicolumn{1}{c|}{Boulby}                & Belmont                                       & \multicolumn{1}{c|}{\textless 9.8}    & \multicolumn{1}{c|}{\textless 0.44}  & \multicolumn{1}{c|}{\textless 0.66}  & \multicolumn{1}{c|}{\textless 0.75}  & \multicolumn{1}{c|}{\textless 0.29}  & \multicolumn{1}{c|}{\textless 1.39}  & \multicolumn{1}{c|}{\textless 2.01}       & \multicolumn{1}{c|}{\textless 0.25}         & \multicolumn{1}{c|}{\textless 0.18}         & \multicolumn{1}{c|}{-}                      & \multicolumn{1}{c|}{\textless 0.10}         \\ \hline
%\multicolumn{1}{|c|}{190705}                  & \multicolumn{1}{c|}{Boulby}                & Merrybent                                     & \multicolumn{1}{c|}{5.9±2.6}          & \multicolumn{1}{c|}{\textless 0.50}  & \multicolumn{1}{c|}{\textless 0.50}  & \multicolumn{1}{c|}{\textless 0.57}  & \multicolumn{1}{c|}{\textless 0.32}  & \multicolumn{1}{c|}{\textless 1.31}  & \multicolumn{1}{c|}{\textless 2.20}       & \multicolumn{1}{c|}{\textless 0.19}         & \multicolumn{1}{c|}{1.6±0.1}                & \multicolumn{1}{c|}{-}                      & \multicolumn{1}{c|}{\textless 0.08}         \\ \hline
\multicolumn{1}{|c|}{\multirow{2}{*}{190706}} & \multicolumn{1}{c|}{Boulby}                & Belmont                                       & \multicolumn{1}{c|}{\textless 9.5}    & \multicolumn{1}{c|}{\textless 0.45}  & \multicolumn{1}{c|}{\textless 0.66}  & \multicolumn{1}{c|}{0.53±0.12}       & \multicolumn{1}{c|}{\textless 0.28}  & \multicolumn{1}{c|}{\textless 1.32}  & \multicolumn{1}{c|}{\textless 2.09}       & \multicolumn{1}{c|}{\textless 0.25}         & \multicolumn{1}{c|}{\textless 0.25}         & \multicolumn{1}{c|}{-}                      & \multicolumn{1}{c|}{\textless 0.13}         \\ \cline{2-14} 
\multicolumn{1}{|c|}{}                        & \multicolumn{1}{c|}{Kamioka}               & Lab-C Ge                                                 & \multicolumn{1}{c|}{\textless 9.4}    & \multicolumn{1}{c|}{\textless 0.69}  & \multicolumn{1}{c|}{\textless 0.50}  & \multicolumn{1}{c|}{\textless 0.86}  & \multicolumn{1}{c|}{\textless 2.26}  & \multicolumn{1}{c|}{\textless 1.10}  & \multicolumn{1}{c|}{\textless 1.9}        & \multicolumn{1}{c|}{\textless 0.29}                      & \multicolumn{1}{c|}{\textless 0.19}         & \multicolumn{1}{c|}{\textless 0.19}                      & \multicolumn{1}{c|}{\textless 0.26}         \\ \hline
%\multicolumn{1}{|c|}{}                        & \multicolumn{1}{c|}{Kamioka}               & Lab-C Ge                                                 & \multicolumn{1}{c|}{\textless 7.3}    & \multicolumn{1}{c|}{\textless 0.64}  & \multicolumn{1}{c|}{\textless 0.39}  & \multicolumn{1}{c|}{\textless 0.59}  & \multicolumn{1}{c|}{\textless 1.76}  & \multicolumn{1}{c|}{\textless 0.83}  & \multicolumn{1}{c|}{\textless 1.7}        & \multicolumn{1}{c|}{-}                      & \multicolumn{1}{c|}{\textless 0.15}         & \multicolumn{1}{c|}{-}                      & \multicolumn{1}{c|}{\textless 0.20}         \\ \hline
\multicolumn{1}{|c|}{190801}                  & \multicolumn{1}{c|}{Canfranc}              & ge-Anayet                                        & \multicolumn{1}{c|}{\textless 14}     & \multicolumn{1}{c|}{0.92}       & \multicolumn{1}{c|}{\textless 1.5}   & \multicolumn{1}{c|}{\textless 0.77}  & \multicolumn{1}{c|}{\textless 0.80}  & \multicolumn{1}{c|}{\textless 1.17}  & \multicolumn{1}{c|}{\textless 1.44}       & \multicolumn{1}{c|}{\textless 0.18}         & \multicolumn{1}{c|}{2.7±0.2}                & \multicolumn{1}{c|}{\textless 0.23}         & \multicolumn{1}{c|}{\textless 0.18}         \\ \hline
%\multicolumn{1}{|c|}{190802}                  & \multicolumn{1}{c|}{Boulby}                & Merrybent                                     & \multicolumn{1}{c|}{\textless 8.44}   & \multicolumn{1}{c|}{\textless 0.57}  & \multicolumn{1}{c|}{\textless 0.56}  & \multicolumn{1}{c|}{\textless 0.68}  & \multicolumn{1}{c|}{\textless 0.48}  & \multicolumn{1}{c|}{\textless 1.18}  & \multicolumn{1}{c|}{\textless 2.54}       & \multicolumn{1}{c|}{\textless 0.17}         & \multicolumn{1}{c|}{4.71±0.20}              & \multicolumn{1}{c|}{-}                      & \multicolumn{1}{c|}{\textless 0.09}         \\ \hline
\multicolumn{1}{|c|}{190803}                  & \multicolumn{1}{c|}{Canfranc}              & ge-Asterix                                       & \multicolumn{1}{c|}{\textless 3.5}      & \multicolumn{1}{c|}{\textless 0.31}  & \multicolumn{1}{c|}{0.39±0.21}       & \multicolumn{1}{c|}{0.55±0.22}       & \multicolumn{1}{c|}{\textless 0.36}  & \multicolumn{1}{c|}{\textless 0.74}  & \multicolumn{1}{c|}{\textless 1.4}        & \multicolumn{1}{c|}{\textless 0.09}         & \multicolumn{1}{c|}{3.5±0.1}                & \multicolumn{1}{c|}{\textless 0.08}         & \multicolumn{1}{c|}{\textless 0.07}         \\ \hline
\multicolumn{1}{|c|}{190804}                  & \multicolumn{1}{c|}{Boulby}                & Belmont                                       & \multicolumn{1}{c|}{\textless 11}     & \multicolumn{1}{c|}{\textless 0.46}  & \multicolumn{1}{c|}{0.67±0.21}       & \multicolumn{1}{c|}{\textless 0.67}  & \multicolumn{1}{c|}{\textless 0.38}  & \multicolumn{1}{c|}{\textless 1.98}  & \multicolumn{1}{c|}{\textless 2.57}       & \multicolumn{1}{c|}{\textless 0.20}         & \multicolumn{1}{c|}{4.60±0.24}              & \multicolumn{1}{c|}{-}                      & \multicolumn{1}{c|}{\textless 0.10}         \\ \hline
\multicolumn{1}{|c|}{190805} & \multicolumn{1}{c|}{Canfranc}              & ge-Oroel                             & \multicolumn{1}{c|}{\textless 4.6}    & \multicolumn{1}{c|}{\textless 0.52}  & \multicolumn{1}{c|}{0.53±0.44}       & \multicolumn{1}{c|}{0.57±0.40}       & \multicolumn{1}{c|}{\textless 0.44}  & \multicolumn{1}{c|}{\textless 0.98}  & \multicolumn{1}{c|}{\textless 1.18}       & \multicolumn{1}{c|}{\textless 0.10}         & \multicolumn{1}{c|}{9.44±0.10}              & \multicolumn{1}{c|}{\textless 0.10}         & \multicolumn{1}{c|}{\textless 0.09}         \\ \hline %\cline{2-14} 
%\multicolumn{1}{|c|}{}                        & \multicolumn{1}{c|}{Kamioka}               & IPMU-P                                           & \multicolumn{1}{c|}{\textless 103}    & \multicolumn{1}{c|}{\textless 1.6}   & \multicolumn{1}{c|}{\textless 3.2}   & \multicolumn{1}{c|}{\textless 4.9}   & \multicolumn{1}{c|}{\textless 16}    & \multicolumn{1}{c|}{\textless 7.0}   & \multicolumn{1}{c|}{\textless 18}         & \multicolumn{1}{c|}{-}                      & \multicolumn{1}{c|}{8.83±0.82}              & \multicolumn{1}{c|}{-}                      & \multicolumn{1}{c|}{\textless 1.2}          \\ \hline
\multicolumn{1}{|c|}{190806} & \multicolumn{1}{c|}{Boulby}                & Merrybent                       & \multicolumn{1}{c|}{\textless{}8.09}  & \multicolumn{1}{c|}{\textless 0.43}  & \multicolumn{1}{c|}{0.49±0.11}       & \multicolumn{1}{c|}{1.27±0.13}       & \multicolumn{1}{c|}{\textless 0.26}  & \multicolumn{1}{c|}{\textless 1.23}  & \multicolumn{1}{c|}{\textless 1.78}       & \multicolumn{1}{c|}{\textless 0.14}         & \multicolumn{1}{c|}{9.35±0.22}              & \multicolumn{1}{c|}{-}                      & \multicolumn{1}{c|}{\textless 0.07}         \\ \hline %\cline{2-14} 
%\multicolumn{1}{|c|}{}                        & \multicolumn{1}{c|}{Kamioka}               & IPMU-N                                           & \multicolumn{1}{c|}{\textless 93}     & \multicolumn{1}{c|}{\textless 3.9}   & \multicolumn{1}{c|}{\textless 3.3}   & \multicolumn{1}{c|}{\textless 2.6}   & \multicolumn{1}{c|}{\textless 19}    & \multicolumn{1}{c|}{\textless 6.4}   & \multicolumn{1}{c|}{\textless 65}         & \multicolumn{1}{c|}{-}                      & \multicolumn{1}{c|}{5.5±0.9}                & \multicolumn{1}{c|}{-}                      & \multicolumn{1}{c|}{\textless 1.4}          \\ \hline
\multicolumn{1}{|c|}{190901} & \multicolumn{1}{c|}{Canfranc}              & ge-Asterix                         & \multicolumn{1}{c|}{\textless 4.3}    & \multicolumn{1}{c|}{\textless 0.30}  & \multicolumn{1}{c|}{0.42±0.27}       & \multicolumn{1}{c|}{0.37±0.27}       & \multicolumn{1}{c|}{\textless 0.46}  & \multicolumn{1}{c|}{\textless 1.20}  & \multicolumn{1}{c|}{\textless 1.47}       & \multicolumn{1}{c|}{\textless 0.15}         & \multicolumn{1}{c|}{4.85±0.12}              & \multicolumn{1}{c|}{\textless 0.10}         & \multicolumn{1}{c|}{\textless 0.13}         \\ \hline %\cline{2-14} 
%\multicolumn{1}{|c|}{}                        & \multicolumn{1}{c|}{Kamioka}               & IPMU-P                                                 & \multicolumn{1}{c|}{\textless 110}    & \multicolumn{1}{c|}{\textless 2.3}   & \multicolumn{1}{c|}{\textless 2.9}   & \multicolumn{1}{c|}{\textless 2.1}   & \multicolumn{1}{c|}{\textless 14.9}  & \multicolumn{1}{c|}{\textless 12.2}  & \multicolumn{1}{c|}{\textless 27}         & \multicolumn{1}{c|}{-}                      & \multicolumn{1}{c|}{5.6±0.7}                & \multicolumn{1}{c|}{-}                      & \multicolumn{1}{c|}{\textless 1.1}          \\ \hline
\multicolumn{1}{|c|}{190902} & \multicolumn{1}{c|}{Boulby}                & Belmont                         & \multicolumn{1}{c|}{\textless 5.52}   & \multicolumn{1}{c|}{\textless 0.26}  & \multicolumn{1}{c|}{0.53±0.10}       & \multicolumn{1}{c|}{0.63±0.09}       & \multicolumn{1}{c|}{\textless 0.33}  & \multicolumn{1}{c|}{\textless 1.22}  & \multicolumn{1}{c|}{\textless 1.32}       & \multicolumn{1}{c|}{\textless 0.10}         & \multicolumn{1}{c|}{8.78±0.18}              & \multicolumn{1}{c|}{-}                      & \multicolumn{1}{c|}{\textless 0.05}         \\ \hline %\cline{2-14} 
%\multicolumn{1}{|c|}{}                        & \multicolumn{1}{c|}{Kamioka}               & IPMU-N                                                 & \multicolumn{1}{c|}{\textless 71}     & \multicolumn{1}{c|}{\textless 4.9}   & \multicolumn{1}{c|}{\textless 3.2}   & \multicolumn{1}{c|}{\textless 2.5}   & \multicolumn{1}{c|}{\textless 19}    & \multicolumn{1}{c|}{\textless 8.0}   & \multicolumn{1}{c|}{\textless 46}         & \multicolumn{1}{c|}{-}                      & \multicolumn{1}{c|}{6.4±0.9}                & \multicolumn{1}{c|}{-}                      & \multicolumn{1}{c|}{\textless 1.4}          \\ \hline
\multicolumn{1}{|c|}{190903}                  & \multicolumn{1}{c|}{Canfranc}               &   ge-Asterix                                               & \multicolumn{1}{c|}{\textless 4.4}     & \multicolumn{1}{c|}{\textless 0.37}   & \multicolumn{1}{c|}{0.59±0.28 }   & \multicolumn{1}{c|}{0.35±0.28}   & \multicolumn{1}{c|}{\textless 0.54}  & \multicolumn{1}{c|}{\textless 1.7}   & \multicolumn{1}{c|}{\textless 1.5}         & \multicolumn{1}{c|}{\textless 0.14}                      & \multicolumn{1}{c|}{4.9±0.1}                & \multicolumn{1}{c|}{\textless 0.10}   & \multicolumn{1}{c|}{\textless 0.09} \\ \hline
\multicolumn{1}{|c|}{\multirow{2}{*}{190905}}                  & \multicolumn{1}{c|}{Kamioka}               & Lab-C Ge                                              & \multicolumn{1}{c|}{\textless 8.6}    & \multicolumn{1}{c|}{\textless 0.21}  & \multicolumn{1}{c|}{0.72±0.20}         & \multicolumn{1}{c|}{0.70±0.16}         & \multicolumn{1}{c|}{\textless 5.2}         & \multicolumn{1}{c|}{\textless 1.1}   & \multicolumn{1}{c|}{\textless 1.57}        & \multicolumn{1}{c|}{\textless 0.09}                      & \multicolumn{1}{c|}{6.6±0.2}                & \multicolumn{1}{c|}{\textless 0.09}                      & \multicolumn{1}{c|}{\textless 0.13}   \\ \cline{2-14} 
\multicolumn{1}{|c|}{}                        & \multicolumn{1}{c|}{Kamioka}               & Lab-C Ge, Ra Disk                                          & \multicolumn{1}{c|}{-}                 & \multicolumn{1}{c|}{\textless 0.29}  & \multicolumn{1}{c|}{0.58±0.25}  & \multicolumn{1}{c|}{\textless 0.39}  & \multicolumn{1}{c|}{-}                & \multicolumn{1}{c|}{-}                & \multicolumn{1}{c|}{-}                     & \multicolumn{1}{c|}{-}                       & \multicolumn{1}{c|}{-}                       & \multicolumn{1}{c|}{-}                       & \multicolumn{1}{c|}{-}                       \\ \hline
%\multicolumn{1}{|c|}{190906}                  & \multicolumn{1}{c|}{Kamioka}               & Lab-C Ge                                                 & \multicolumn{1}{c|}{\textless{}7.0}   & \multicolumn{1}{c|}{\textless{}0.19} & \multicolumn{1}{c|}{1.09±0.23}       & \multicolumn{1}{c|}{0.45±0.14}       & \multicolumn{1}{c|}{6.0±2.0}         & \multicolumn{1}{c|}{\textless{}0.53} & \multicolumn{1}{c|}{\textless{}1.1}       & \multicolumn{1}{c|}{-}                       & \multicolumn{1}{c|}{5.92±0.21}              & \multicolumn{1}{c|}{-}                       & \multicolumn{1}{c|}{\textless{}0.10}        \\ \hline
%\multicolumn{1}{|c|}{191001}                  & \multicolumn{1}{c|}{Kamioka}               & Lab-C Ge                                                 & \multicolumn{1}{c|}{\textless 5.2}    & \multicolumn{1}{c|}{\textless 0.26}  & \multicolumn{1}{c|}{1.62±0.24}       & \multicolumn{1}{c|}{0.55±0.13}       & \multicolumn{1}{c|}{4.6±1.6}         & \multicolumn{1}{c|}{\textless 0.45}  & \multicolumn{1}{c|}{\textless 1.13}       & \multicolumn{1}{c|}{-}                       & \multicolumn{1}{c|}{5.57±0.17}            & \multicolumn{1}{c|}{-}                       & \multicolumn{1}{c|}{0.13±0.08}              \\ \hline
\multicolumn{1}{|c|}{200101}                  & \multicolumn{1}{c|}{Kamioka}               & IPMU-N                                                 & \multicolumn{1}{c|}{\textless 6.80}     & \multicolumn{1}{c|}{\textless 0.35}   & \multicolumn{1}{c|}{0.98±0.18}   & \multicolumn{1}{c|}{1.00±0.15}   & \multicolumn{1}{c|}{8.24±1.68}    & \multicolumn{1}{c|}{\textless 0.54}   & \multicolumn{1}{c|}{\textless 0.95}         & \multicolumn{1}{c|}{\textless 0.08}                 & \multicolumn{1}{c|}{6.25±0.17}                & \multicolumn{1}{c|}{\textless 0.18}                      & \multicolumn{1}{c|}{\textless 0.13}          \\ \hline
%\multicolumn{1}{|c|}{200102}                  & \multicolumn{1}{c|}{Kamioka}               & IPMU-P                                                 & \multicolumn{1}{c|}{\textless 122}    & \multicolumn{1}{c|}{\textless 2.5}   & \multicolumn{1}{c|}{\textless 3.1}   & \multicolumn{1}{c|}{\textless 3.3}   & \multicolumn{1}{c|}{\textless 16}    & \multicolumn{1}{c|}{\textless 7.9}   & \multicolumn{1}{c|}{\textless 25}         & \multicolumn{1}{c|}{-}                      & \multicolumn{1}{c|}{7.0±0.8}                & \multicolumn{1}{c|}{-}                      & \multicolumn{1}{c|}{\textless 0.98}         \\ \hline
\multicolumn{1}{|c|}{200103}                  & \multicolumn{1}{c|}{Kamioka}               & IPMU-N                                                 & \multicolumn{1}{c|}{\textless 8.46}     & \multicolumn{1}{c|}{0.51±0.12}   & \multicolumn{1}{c|}{1.42±0.25}   & \multicolumn{1}{c|}{0.84±0.17}   & \multicolumn{1}{c|}{\textless 2.11}    & \multicolumn{1}{c|}{\textless 0.88}   & \multicolumn{1}{c|}{\textless 1.43}         & \multicolumn{1}{c|}{\textless 0.12}                 & \multicolumn{1}{c|}{0.18±0.07}                & \multicolumn{1}{c|}{\textless 0.13}                      & \multicolumn{1}{c|}{\textless 0.16}          \\ \hline

\multicolumn{1}{|c|}{200104}                  & \multicolumn{1}{c|}{Kamioka}               & IPMU-N                                                 & \multicolumn{1}{c|}{\textless 8.39}     & \multicolumn{1}{c|}{\textless 0.36}   & \multicolumn{1}{c|}{1.48±0.24}   & \multicolumn{1}{c|}{0.84±0.18}   & \multicolumn{1}{c|}{\textless 3.45}    & \multicolumn{1}{c|}{\textless 0.95}   & \multicolumn{1}{c|}{\textless 1.02}         & \multicolumn{1}{c|}{\textless 0.08}                 & \multicolumn{1}{c|}{\textless 0.28}                & \multicolumn{1}{c|}{\textless 0.23}                      & \multicolumn{1}{c|}{\textless 0.11}          \\ \hline

%\multicolumn{3}{c|}{}                                                                                                                         & \multicolumn{1}{c|}{238U   up}        & \multicolumn{1}{c|}{238U   lo}       & \multicolumn{1}{c|}{232Th   up}      & \multicolumn{1}{c|}{232Th   lo}      & \multicolumn{1}{c|}{235U   up}       & \multicolumn{1}{c|}{235U   lo}       & \multicolumn{1}{c|}{40K}                  & \multicolumn{1}{c|}{138La}                  & \multicolumn{1}{c|}{176Lu}                  & \multicolumn{1}{c|}{134Cs}                  & \multicolumn{1}{c|}{137Cs}                  \\ \cline{4-14}
\end{tabular}
\caption{Summary of \GdSOw\ assay results by HPGe detectors. The sample identifier is coded as follows: YYMM\#\# where YYMM is the year and month of production and \#\# refers to the batch number produced within that month. The measurements of each radioactive chain are separated into the early part of the chain (E) and the late part of the chain (L). The isotopes identified are the longest lived within each sub-chain, and the activities are estimated assuming secular equilibrium (eq.) within each sub-chain.
\label{Tab:GeSummary}}
\end{table}
\end{landscape}

\begin{table}
\centering
\begin{tabular}{|c|c|c|cc|cc|c|}
\hline
\multirow{2}{*}{Chain} & \multirow{2}{*}{\begin{tabular}[c]{@{}c@{}}Part of \\ Chain\end{tabular}} & \multirow{2}{*}{$t_{1/2}$} & \multicolumn{2}{c|}{SK-Gd Req.} & \multicolumn{2}{c|}{HPGe} & ICP-MS \\ \cline{4-8} 
 &  &  & \multicolumn{1}{c|}{\begin{tabular}[c]{@{}c@{}}Specific\\ Activity\\ (mBq/kg)\end{tabular}} & \begin{tabular}[c]{@{}c@{}}Total\\ Budget\\ (Bq)\end{tabular} & \multicolumn{1}{c|}{\begin{tabular}[c]{@{}c@{}}FM \\ (Bq)\end{tabular}} & \begin{tabular}[c]{@{}c@{}}UL\\ (Bq)\end{tabular} & \begin{tabular}[c]{@{}c@{}}Total \\ (Bq) \end{tabular} \\ \hline\hline
\multirow{2}{*}{$^{238}$U} & RI $^{238}$U & 4.5 Gy & \multicolumn{1}{c|}{$<5$} & 650 & \multicolumn{1}{c|}{--} & -- & $0.34 \pm 0.15$ \\
 &E, $^{238}$U eq.& 4.5 Gy & \multicolumn{1}{c|}{$<5$} & 650 & \multicolumn{1}{c|}{$0$} & $<89$ & -- \\
 &L, $^{226}$Ra eq.& 1602 y & \multicolumn{1}{c|}{$<0.5$} & 65 & \multicolumn{1}{c|}{$0.25 \pm 0.06$} & $<5.2$ & -- \\ \hline
\multirow{3}{*}{$^{232}$Th} & RI $^{232}$Th & 14 Gy & \multicolumn{1}{c|}{$<0.05$} & 6.5 & \multicolumn{1}{c|}{--} & -- & $0.25 \pm 0.07$ \\
 &E, $^{228}$Ra eq.& 5.7 y & \multicolumn{1}{c|}{$<0.05$} & 6.5 & \multicolumn{1}{c|}{$4.1 \pm 0.4$} & $<11$ & -- \\
 &L, $^{228}$Th eq.& 1.9 y & \multicolumn{1}{c|}{$<0.05$} & 6.5 & \multicolumn{1}{c|}{$3.8 \pm 0.3$} & $<8.9$ & -- \\ \hline
\multirow{2}{*}{$^{235}$U} &E, $^{235}$U eq.& 0.7 Gy & \multicolumn{1}{c|}{$<30$} & 3900 & \multicolumn{1}{c|}{$4.1 \pm 0.8$} & $<15$ & -- \\
 &L, $^{227}$Ac eq.& 21.7 y & \multicolumn{1}{c|}{$<30$} & 3900 & \multicolumn{1}{c|}{$3.3 \pm 0.7$} & $<19$ & -- \\ \hline
\end{tabular}
    \caption{Summary of HPGe and ICP-MS measurement results for all samples of \GdSOw\ assayed and approved for dissolving into SK, compared with the total SK-Gd radioactivity budget assuming 0.2\% loading of 130 tons of \GdSOw. The measurements of each radioactive chain are separated into those for the parent radioactive isotopes (RIs), the early part of the chain (E) and the late part of the chain (L). The HPGe assay results are combined in two ways to give an estimate of the minimum and maximum total added radioactivity to SK. See the text for an explanation of the methods for combining finite measurements (FM) and upper limits (UL).}
    \label{tab:resultssummary}
\end{table}

\section{New purification process to reduce Radium} \label{sec:new_purification}% Ikeda

To reduce the radium concentration in future batches of \GdSOw\ for SK, a new process has been developed. It utilizes the difference of precipitation pH between Gd hydroxide and Ra hydroxide. By adding a base to an acidic aqueous solution of the raw material and adjusting the pH of the aqueous solution between 8 and 9, Gd precipitates as Gd hydroxide while maintaining the state in which Ra is still dissolved in the aqueous phase. As bases for pH adjustment, ammonia, sodium hydroxide, and potassium hydroxide can be considered. It is most preferable to use ammonia for pH adjustment to avoid contamination with other metals such as sodium or potassium. During a test, 25\% by mass aqueous ammonia was gradually added to the aqueous Gd chloride solution after dissolving Gd oxide to adjust the pH to 8.0. Then Gd hydroxide can be obtained as a precipitate. The amount of ammonia water added to 10~kg of the Gd chloride solution was 4.7~kg. The precipitate was separated by filtration and washed with water until the chloride ion concentration of the filtrate became 1000~mg/L or less to obtain gadolinium hydroxide. The obtained hydroxide is again dissolved in 35\% by mass hydrochloric acid aqueous solution to obtain an acidic aqueous Gd solution which is an input to the solvent extraction. As shown in Table~\ref{tab:NewProc}, we succeeded in reducing Ra to less than 1\% of its previous value with this new process.

\begin{table}
\begin{tabular}{ccc}

                              & Only solvent extraction & With hydroxide precipitation \\ \hline \hline
pH during precipitation       & -                       & 8.0                            \\ \hline
pH during solvent extraction  & 1.1                     & 1.1                          \\ \hline
Gd recovery rate {[}\%{]}     & 91                      & 90                           \\ \hline
$^{238}$U    by ICP-MS {[}ppb{]}   & 16                      & \textless 0.8                \\ \hline
$^{232}$Th     by ICP-MS {[}ppb{]} & 1.2                     & \textless 0.02               \\ \hline
$^{228}$Ra by HPGe {[}mBq/kg{]}      & 162                     & \textless 0.4                \\ \hline
\end{tabular}
\caption{Relevant chemical properties during Gd processing with and without the new purification procedures.}
\label{tab:NewProc}
\end{table}

\section{Conclusions} \label{sec:conclusions} %% Sekiya
This paper reports the development and detailed properties of about 13 tons of \GdSOw\ which has been dissolved into SK in the summer of 2020. We evaluated the impact of radioactive impurities in \GdSOw\ on DSNB searches and solar neutrino observations. We confirmed the need to reduce radioactive and fluorescent impurities by about three orders of magnitude from commercially available high-purity \GdSOw. Therefore, we developed a method to remove impurities from \GdOx\ by acid dissolution, solvent extraction, and pH control processes, and then we developed a high-purity sulfation process. All of the produced high-purity \GdSOw\ batches were sampled and assayed by ICP-MS and HPGe detectors to evaluate their quality. Because the HPGe assays require long measurement times to achieve sufficient sensitivity to distinguish very low levels of radioactive impurities, measurements were performed in cooperation with HPGe detectors at LSC in Spain, Boulby in the UK, and Kamioka in Japan.

In the first half of production, the assayed batch purities were consistent with the specifications. However, it was found that the \GdSOw\ in the latter half of the produced batches contained one order of magnitude more $^{228}$Ra than the budgeted mean contamination. This was correlated with the corresponding characteristics of the raw material \GdOx, in which an intrinsically large contamination was present and could not affordably be replaced with raw material of ideal condition. However, since the increase of the background event rate for solar neutrino observations due to this activity of $^{228}$Ra is at the same level as the background in the pure water phase of SK, the decision was made to introduce these later batches into SK anyway. 

A new method to remove $^{228}$Ra from \GdOx\ has been subsequently established. 
Introduced into SK in 2022, the latest 26 tons of \GdSOw\ show the required high-purity.

\section*{Acknowledgments}
We gratefully acknowledge the Super-Kamiokande experiment. 
It has been built and operated from funding by the Japanese Ministry of Education, Culture, Sports, Science and Technology, the U.S. Department of Energy, and the U.S. National Science Foundation. 
We also acknowledge the cooperation of the Kamioka Mining
and Smelting Company, Japan.  
Some of us have been supported by funds from 
the Ministry of Education, Japan 2018R1D1A3B07050696, 2018R1D1A1B07049158, 
 the Japan Society for the Promotion of Science (JSPS) KAKENHI Grants Grant-in-Aid for Scientific Research on Innovative Areas No. 26104004, 26104006, 17H06365, 19H05807, 19H05808, Grant-in-Aid for Specially Promoted Research No. 26000003, Grant-in-Aid for Young Scientists No. 17K14290, and Grant-in-Aid for JSPS Research Fellow No. 18J00049, Universities and Innovation grant PID2021-124050NB-C31, the European Union’s Horizon 2020 Research and Innovation Programme under the Marie Sklodowska-Curie 2020-MSCA-RISE-2019 SK2HK grant agreement no. 872549.
We acknowledge Israel Chemicals Ltd UK (ICL-UK) for hosting the U.K. Science and Technology Facilities Council's (STFC) Boulby Underground Laboratory. The authors from Sheffield were supported by the STFC under award numbers ST/R000069/1, ST/T00200X/1, ST/V002821/1, ST/V006185/1, and ST/S000747/1.
We are grateful to the Laboratorio Subterráneo de Canfranc (Spain) for supporting the low-background materials screening work.

\bibliographystyle{ptephy}
\bibliography{Reference}

\appendix
\section{X-ray diffraction analysis}\label{appendix:xraydiff}
\begin{figure}[htbp]
    \centering 
    \includegraphics[width=\textwidth]
    {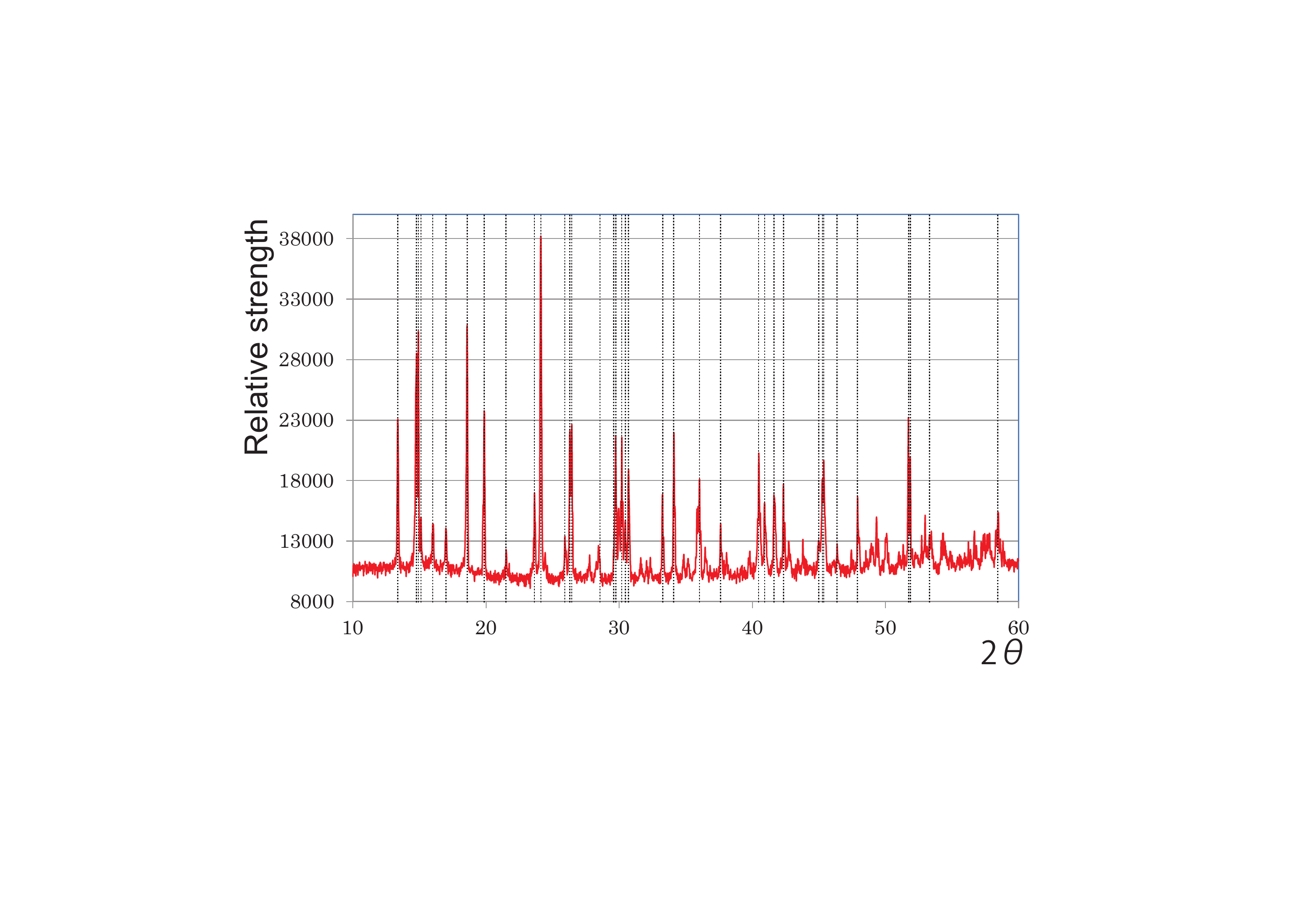}
    \caption{\label{fig:xray}
    One example of x-ray diffraction data (red); peak positions in the ICPDS database [\# 01-81-1794: gadolinium sulfate octahydrate] are shown as dashed lines (black).}
\end{figure}

An example of x-ray diffraction data of one batch sample is shown in Figure~\ref{fig:xray}. Since \GdSOw\ particles have high crystallinity, the orientation dependence of each peak intensity is high. Therefore, the intensity may change depending on how it is packed in a holder, so the peak positions (not the relative intensities) must be checked. As a result, all the peak positions are in agreement with the X-ray diffraction database :ICPDS \# 01-81-1794: Gadolinium Sulfate octahydrate.

\section{Thermogravimetric analysis}\label{appendix:thermograv}
\begin{figure}[htbp]
    \centering 
    \includegraphics[width=\textwidth]
    %{img/KamGeSpectra.eps}
    {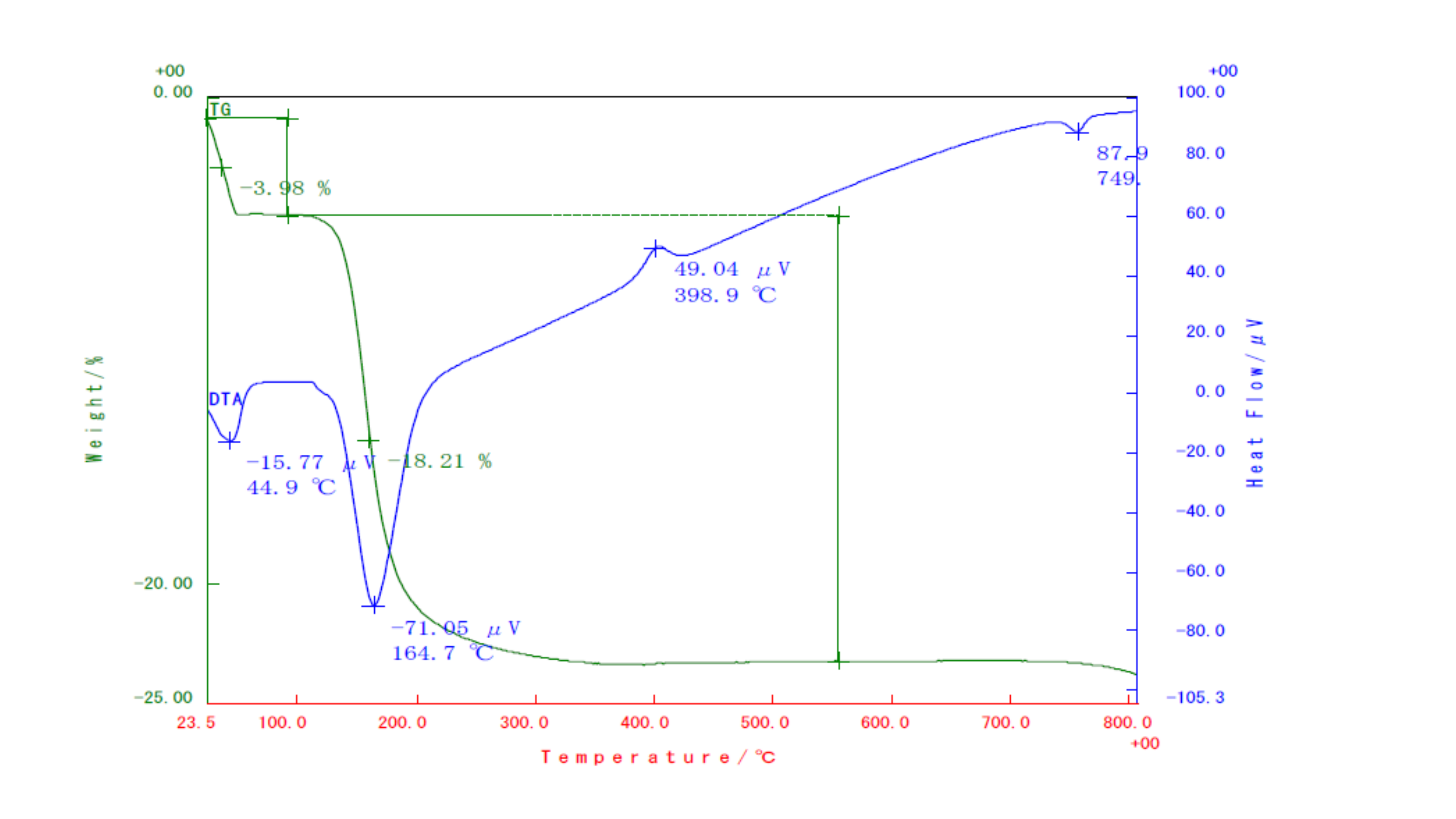}
    \caption{\label{fig:tgdta}
    The mass loss and the heat flow as functions of sample temperature. 
    }
\end{figure}

An example of thermogravimetric analysis data of one batch sample is shown in Figure~\ref{fig:tgdta}. The sample is heated from room temperature to 800$^\circ$ C, and the mass change depending on the temperature is measured. Figure~\ref{fig:tgdta} shows the mass loss and the heat flow as functions of sample temperature. The mass loss observed near 100$^\circ$ C is mainly due to the evaporation of the water remaining during purification, and the mass loss at higher temperatures is due to the evaporation of the octahydrate from \GdSOw. The expected mass loss due to going from an  octahydrate to an anhydrous form of gadolinium sulfate is 19.3\%.

\end{document}